\documentclass[aps,prl,superscriptaddress,twocolumn,floatfix]{revtex4-2}
\usepackage[abs]{overpic}
\usepackage{graphicx,graphics}
\usepackage{dcolumn}
\usepackage{amsmath,amssymb,amsfonts}
\usepackage{wasysym}
\usepackage{latexsym,verbatim}
\usepackage{bm}
\usepackage[dvipsnames]{xcolor}
\usepackage[framemethod=default]{mdframed}
\usepackage[breaklinks=true,colorlinks,citecolor=blue,linkcolor=blue,urlcolor=blue]{hyperref}
\usepackage[makeroom]{cancel}
\usepackage{fancybox}
\usepackage{ulem}

\newcommand{\be}{\begin{eqnarray}}
\newcommand{\ee}{\end{eqnarray}}
\newcommand{\nn}{\nonumber\\}

\begin{document}

\title{Broken Lorentz symmetry and violation of the Wiedemann-Franz law in topological insulators}

\author{Feng Liu}
\email{feng.liu-6@postgrad.manchester.ac.uk}
\author{A. Daria Dumitriu-I.}
\email{alexandra-daria.dumitriu-iovanescu@outlook.com}
\author{Alessandro Principi}%
\email{alessandro.principi@manchester.ac.uk}
\affiliation{%
Department of Physics and Astronomy, University of Manchester, M13 9PL Manchester (UK)
 }%


\begin{abstract}
We study a two-dimensional topological insulator in the presence of a static non-uniform gravitational field, which mimics the variations in the temperature distribution. 
We derive an effective boundary free energy functional for the gravitational field and show that, in contrast to the case of massive Dirac fermions, the addition of a Newtonian mass term significantly modifies the quantum anomalous behavior of the system. A non-zero bulk thermal current appears, which violates the Wiedemann-Franz law.
The systematic approach we develop to calculate the contribution of edge states to thermal Hall conductivity and energy magnetization can easily be extended to other models.

\end{abstract}

\maketitle

%
{\it Introduction}---The thermal Hall effect has the potential to reveal topological properties that are inaccessible through the conventional (electrical) Hall effect. In fact, it enables probing invariants in systems with broken $U(1)$ symmetry~\cite{golan2018probing}, with neutral excitations~\cite{grissonnanche2020chiral}, and uncovering complex boundary state structures in strongly correlated systems~\cite{kane1997quantized}.
In the thermal Hall effect, heat currents are deflected perpendicular to the temperature gradient. The phenomenon is analogous to the electrical Hall effect but with heat flow in place of charge currents. 

Experimental evidence for the thermal Hall effect has been steadily growing in various systems, including multiferroics~\cite{ideue2017giant}, topological insulators~\cite{shimizu2015quantum}, quantum magnets~\cite{zhang2024thermal}, and various other materials~\cite{grissonnanche2019giant,samajdar2019enhanced,fulga2020temperature,grissonnanche2020chiral}.  
Significant theoretical effort has been put into the microscopic description and study of this phenomenon~\cite{cooper1997thermoelectric,qin2011energy,luttinger1964theory,shitade2014heat,paul2003thermal,cappelli2002thermal,bradlyn2015low,stone2012gravitational,nomura2012cross}. 
From a microscopic perspective~\cite{luttinger1964theory,qin2011energy,cooper1997thermoelectric}, thermal responses can be calculated using Luttinger's “trick”~\cite{luttinger1964theory}. This exploits the equivalence in the linear-response regime between a non-flat metric tensor (also termed  ``gravitational potential'' in analogy to the electric one) and temperature gradients that generate energy currents.
From an effective theory perspective~\cite{stone2012gravitational,bradlyn2015low,cappelli2002thermal,read2000paired} the thermal Hall conductivity $\kappa_{xy}$ of a topological insulator at low temperatures is determined by the central charge of the corresponding edge theory~\cite{kane1997quantized}.

Both approaches have limitations that make them not fully satisfactory. Within the microscopic perspective, scaling relations for heat currents have often been assumed without providing any microscopic justification. Moreover, ambiguities remain about how to extend these methods to interacting systems~\cite{qin2011energy,shitade2014heat}.
%
On the other hand, the effective action approach describes edge states by means of a conformal field theory, based on an underlying low-energy topological field theory (\textit{i.e.} the gravitational Chern-Simons theory~\cite{stone2012gravitational,bradlyn2015low}). However, deriving the gravitational Chern-Simons theory directly from a microscopic bulk Hamiltonian poses significant challenges~\cite{sumiyoshi2013quantum}. Therefore, it would be preferable to derive the conformal edge theory (and its associated thermal Hall conductivity) directly from the bulk microscopic Hamiltonian, without relying on any topological field theory as an intermediate step.

In this paper, we carry out this precise task. We derive a long-wavelength description of edge states from a bulk microscopic Hamiltonian without assuming Lorentz symmetry, which is in general broken in a topological insulator.
As a consequence, the energy current cannot be simply defined as the functional derivative of the Hamiltonian with respect to the vielbein field. To resolve this, we introduce an additional source term in the Hamiltonian.
Because this term vanishes when the metric reduces to Luttinger’s one, this ``extended Hamiltonian'' can be used to define the partition function.
Compared to previous approaches~\cite{nakai2016finite,liu2024no}, this method allows one to treat systems that are not massive Dirac fermions.

From the study of a (2+1)-dimensional topological insulator model, we demonstrate that the broken Lorentz symmetry significantly modifies the quantum anomalous behavior of edge states. 
In contrast to massive Dirac fermions~\cite{liu2024no}, in a generic topological insulator the presence of a Newtonian mass term~\cite{qi2011topological} breaks energy conservation when the metric is allowed to vary in space. This suggests that a bulk thermal Hall current can exist even when the gradient of the gravitational field is uniform. This result cannot be obtained from the gravitational Chern-Simons action, which can only be derived from a (2+1)-dimensional massive Dirac fermion model. In the latter model, the bulk thermal Hall current is proportional to higher derivatives of the metric.
Furthermore, we find that the thermal Hall conductivity contains a term proportional to $T^3$ which cannot be obtained from the gravitational Chern-Simons theory.
We conclude that the Wiedemann-Franz law only holds in the low-temperature limit.

\begin{figure}[t!]
    \centering
    \includegraphics{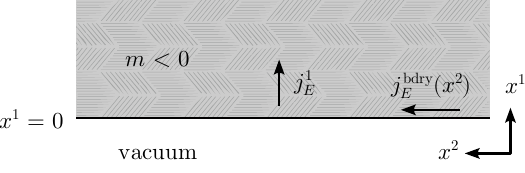}
    \caption{The boundary located at $x^1=0$ separates a (2+1)-dimensional topological insulator system with $m<0$ from the vacuum. A boundary current $j^{\text{bdry}}_E$ flows along the edge, {\it i.e.} in the $x^{2}$ direction, and bulk currents $j^{1}_E$ flow in the direction orthogonal to the the boundary ($x^1$). The textured background represents a non-zero gravitational potential.
    The equilibrium temperature of the edge mode matches that of the upstream heat bath~\cite{czajka2022exotic,vinkler2018approximately}. Throughout the paper, the
    temperature $T$ is treated as uniform in both space and time. Any spatial variation in temperature is interpreted as a variation in gravitational potential~\cite{luttinger1964theory,cooper1997thermoelectric}.
    } 
    \label{fig:one}
\end{figure}

{\it The model}---We consider the system presented in Fig.~\ref{fig:one}, {\it i.e.} a (1+1)-dimensional boundary located at $x^1=0$ and oriented along the direction $x^2$ which separates a (2+1)-dimensional topological insulator system with relativistic mass $m<0$ from the vacuum. The Newtonian mass $m_N$ is assumed to be positive.
The action for the (2+1)-dimensional topological insulator coupled to a gravitational field is (hereafter we set both the particles' speed and $\hbar$ to one)~\cite{qi2011topological,gromov2015thermal}
\be \label{eq:action_def}
S=\int_{x,t} \sqrt{g}\Big[&&\frac{i}{2}(\bar{\psi}e^{\mu}_{\phantom{.}\alpha}\gamma^{\alpha}\nabla_{\mu}\psi-(\nabla_{\mu}\bar{\psi})\gamma^{\alpha}e^{\mu}_{\phantom{.}\alpha}\psi)
\nn
&&-m\bar{\psi}\psi+\frac{1}{m_N}\eta^{ab}e^{\mu}_{\phantom{.}a}e^{\nu}_{\phantom{.}b}(\nabla_{\mu}\bar{\psi})(\nabla_{\nu}\psi)
\Big]
\ee
where we used the short-hand notation $\int_{x,t}=\int d^2x dt$. $\overrightarrow{\nabla}_{\mu}$ ($\overleftarrow{\nabla}_{\mu}$) is a covariant derivative that acts on the right (left) two-component spinor field $\psi$, and it is explicitly given by $\overrightarrow{\nabla}_{\mu}\psi = \overrightarrow{\partial}_{\mu}\psi +[\gamma_{\alpha},\gamma_{\beta}]\omega_{\mu}^{\phantom{.}\alpha\beta} \psi/8$. Here, $\overrightarrow{\partial}_\mu$ denotes the derivative over the temporal ($\mu=0$) and spatial ($\mu = x,y$) directions.
Finally, the combinations  $\gamma^0\gamma^1$, $\gamma^0\gamma^2$ and $\gamma^0$ correspond to the usual Pauli matrices $\sigma_x$, $\sigma_y$ and $\sigma_z$, respectively.

In  Eq.~(\ref{eq:action_def}), we have introduced the metric $g_{\mu\nu}$,  whose determinant, in modulus, is $g$.
The factor $\sqrt{g}$ ensures invariance of the action under changes of coordinates. Throughout this paper, we use the Greek indices $\mu,\nu=0,1,2$ and $\alpha,\beta,\ldots =\hat{0},\hat{1},\hat{2}$ to denote the environment and locally flat (or internal) coordinates, respectively. In what follows, when we refer to space-like directions only, we will use the Latin letters $i,j=1,2$ for the environment coordinates, and $a,b=\hat{1},\hat{2}$ for the internal coordinates. The Minkowski metric in the locally-flat space-time is taken to be $\eta_{\alpha\beta}=\text{diag}(+1,-1,-1)$. The environment and flat metrics, $g_{\mu\nu}$ and $\eta_{\alpha\beta}$, respectively, are related by a vielbein field $e^{\phantom{.}\alpha}_{\mu}$ according to the identity~\cite{carroll2019spacetime}
$g_{\mu\nu}=e^{\phantom{.}\alpha}_{\mu}e^{\phantom{.}\beta}_{\nu}\eta_{\alpha\beta}$. 
Note that the product $\eta^{ab}e^{\mu}_{\phantom{.}a}e^{\nu}_{\phantom{.}b}$ that appears in Eq.~(\ref{eq:action_def}) does not coincide with $g^{\mu\nu}$, because $a,b\neq {\hat 0}$. Thus, such term breaks Lorentz symmetry.

We assume the perturbation to be of the Luttinger's type~\cite{luttinger1964theory}, 
{\it i.e.}
$
e^{\phantom{.}\hat{0}}_{\mu}=\delta^{\phantom{.}\hat{0}}_{\mu}(1+\phi)$, and $e^{\phantom{.}a}_{\mu}=\delta^{\phantom{.}a}_{\mu}$~\cite{bradlyn2015low}.
In what follows, except when explicitly stated otherwise, we set $m_N = 1$.
From Eq.~(\ref{eq:action_def}), the energy-current  is obtained as
$j_E^i=
-e^{\phantom{.}\alpha}_{0} (\delta S/ \delta e^{\phantom{.}\alpha}_{i})$~\cite{gromov2015thermal,bradlyn2015low}.
Similarly, the Hamiltonian acting in an effectively locally-flat space-time is 
\be \label{eq:Ham_full}
H&=&\int_x \sqrt{g}\psi^{\dagger}\gamma^{\hat{0}}\Big[\frac{i}{2}e^{0}_{\phantom{.}\alpha}\gamma^\alpha\omega_0+\frac{i}{2}\omega_0\gamma^\alpha e^{0}_{\phantom{.}\alpha}-\frac{i}{2}e^{j}_{\phantom{.}\alpha}\gamma^{\alpha}\overrightarrow{\nabla}_j
\nn
&&+\frac{i}{2}\overleftarrow{\nabla}_{j}\gamma^{\alpha}e_{\phantom{.}\alpha}^{j}+m-\eta^{ab}e^{\mu}_{\phantom{.}a}e^{\nu}_{\phantom{.}b}\overleftarrow{\nabla}_{\mu}\overrightarrow{\nabla}_{\nu}\Big]\psi
\ee
Because the Hamiltonian breaks Lorentz symmetry, the energy-momentum tensor is no longer symmetric, and thus the energy current cannot be defined as the functional derivative of the Hamiltonian. To address this issue, we introduce the extended Hamiltonian as $H_{\rm ex}=H+H_{\rm src}$, where
\be \label{eq:Ham_ex}
H_{\rm src}&=&-
\int_x \sqrt{g}e^{0}_{\phantom{.}\hat{0}}e^i_{\phantom{.}\hat{0}}\psi^{\dagger}\gamma^{\hat{0}}\Big[\overleftarrow{\partial}_i\overrightarrow{\partial}_0+\overleftarrow{\partial}_0\overrightarrow{\partial}_i
\Big]\psi.
\ee
The time derivatives in Eq.~(\ref{eq:Ham_ex}) should be intended as replaced by the equation of motion generated by the Hamiltonian $H$.
The so-defined Hamiltonian $H_{\rm ex}$ reproduces the energy current obtained from the original action $S$ and the equation of motion obtained from $H$.
Furthermore, the term $H_{\rm src}$ partially restores Lorentz symmetry and allows one to define the energy current as a functional derivative of the Hamiltonian~\cite{supplemental_material}.
%
For a metric corresponding to Luttinger's gravitational potential and in the absence of Lorentz symmetry, we then find the energy current
\be \label{eq:energy_current}
j^2_E
&=&-(1+\phi)^2\frac{\delta H_{\rm ex}}{\delta e^{\phantom{.}\hat{2}}_0}-(1+\phi)\partial_i\Big(\frac{\delta H_{\rm ex}}{\delta \omega_{i\hat{0}\hat{2}}}
\Big).
\ee
This equation is the first fundamental result of our paper~\cite{supplemental_material}. In what follows, we express the energy current as a functional derivative of the extended Hamiltonian $H_{\rm ex}$. In the case of Luttinger's gravitational potential, Eq.~(\ref{eq:energy_current}) holds at the operator level. 

Since the source term $H_{\rm src}$ in Eq.~(\ref{eq:Ham_ex}) is zero when the metric is of the Luttinger's type, the extended Hamiltonian $H_{\rm ex}$ can be used to define the partition function. Thus, the expectation value of the energy current at finite temperature is given by~\cite{supplemental_material}
\be \label{eq:energy_current_expect}
j_E^{2}&=&-(1+\phi)^2\frac{\delta F_{\rm ex}}{\delta e^{\phantom{.}\hat{2}}_{0}}-(1+\phi)\partial_i\Big(\frac{\delta F_{\rm ex}}{\delta \omega_{i\hat{0}\hat{2}}}\Big).
\ee
As usual, we define the free energy $F_{\rm ex} = -\beta^{-1} \ln Z_{\rm ex}$, where $Z_{\rm ex} = \text{Tr}\left(e^{-\beta H_{\rm ex}}\right)$ is the partition function and $\text{Tr}(\ldots)$ represents the trace in the Fock space~\cite{stoof2009ultracold}.

It is important to emphasize that the functional derivative in Eq.~(\ref{eq:energy_current_expect}) is taken assuming the spin connection to be completely general, with all its components different from zero. However, after differentiation, we consider the special case of Luttinger gravity, where Lorentz symmetry is absent. This leads to a vanishing spin connection~\cite{bradlyn2015low,gromov2015thermal}.

{\it Boundary theory}---We derive the Hamiltonian for boundary fermions using the standard procedure from~\cite{nakai2016finite}. Further details can be found in~\cite{supplemental_material}.
We define $e^{\phantom{.}\alpha}_{\mu}=\delta^{\phantom{.}\alpha}_{\mu}+h^{\phantom{.}\alpha}_{\mu}/2$, $e_{\phantom{.}\alpha}^{\mu}=\delta_{\phantom{.}\alpha}^{\mu}-h_{\phantom{.}\alpha}^{\mu}/2$ and $\sqrt{g}=1+h/2$, where $h$, $h^{*\alpha}_{\mu}$ and $h^{\mu}_{*\alpha}$ are generic deviations of the vielbein field and its determinant from their value in a flat spacetime.
We assume the deviations $h$, $h^{\phantom{.}\alpha}_{\mu}$ and $h^{\mu}_{\phantom{.}\alpha}$ to be small.
Therefore, the unperturbed Hamiltonian is
\be
H_0&=&\int d^2x\psi^{\dagger}\Big[-\frac{i}{2}\gamma^{\hat{0}}\gamma^{j}\overrightarrow{\partial}_j+\frac{i}{2}\gamma^{\hat{0}}\overleftarrow{\partial}_{j}\gamma^{j}
\nn
&&+m\gamma^{\hat{0}}-\gamma^{\hat{0}}\eta^{ab}\overleftarrow{\partial}_{a}\overrightarrow{\partial}_{b}
\Big]\psi,
\ee
while other parts in $H_{\rm ex}$ are treated as perturbations.

We assume that close to the boundary, the metric only depends on $x^2$, such that the two directions $x^1$ and $x^2$ completely decouple in the boundary Hamiltonian. Thus, the wave function of the boundary mode obtained from the Hamiltonian $H_0$ factorizes as $\psi(x^1,x^2)=\psi_2(x^2)\psi_1(x^1)$, where $\psi_2(x^2)$ is a plane wave in the $x^2$ direction and $\psi_1(x^1)$ is a two-component evanescent spinor wave function in the $x^1$ direction. The formal expression of the latter is given by~\cite{qi2011topological,supplemental_material}
$\psi_1(x^1)=(e^{-\lambda_1x^1}-e^{-\lambda_2x^1})\vert s\rangle$ with $\lambda_{1,2}=(1\pm \sqrt{1+4m})/2$, and the two component spinor $\vert s\rangle$, corresponding to the edge bound states, satisfies $i\gamma^{\hat{1}}\vert s\rangle =\vert s\rangle$. 
Therefore, the one-body boundary Hamiltonian obtained from the unperturbed bulk Hamiltonian $H_0$ is $\mathcal{H}_0=(i/2)(\overrightarrow{\partial}_2-\overleftarrow{\partial}_2)$.

The derivation of the extended {\it boundary} Hamiltonian $\mathcal{H}_{\rm ex}$ is more involved. We therefore relegate it to the Supplemental Material~\cite{supplemental_material} and quote only the final result
\be \label{eq:Ham_bdry_e}
\mathcal{H}_{\rm ex}&=&\frac{i}{2}(\overrightarrow{\partial}_2-\overleftarrow{\partial}_2)+\frac{i}{2}\zeta(x^2)(\overrightarrow{\partial}_2-\overleftarrow{\partial}_2)
\nn
&&+i\chi(x^2)(-\overleftarrow{\partial}_2\overrightarrow{\partial}_2^2+\overleftarrow{\partial}^2_2\overrightarrow{\partial}_2),
\ee
where
\be
\zeta(x^2)&=&\frac{h}{2}+\Big(1+\frac{h}{2}\Big)\Big((h^{2}_{\phantom{.}\hat{0}}/2)-(h^{2}_{\phantom{.}\hat{2}}/2)
\Big),
\nn
\chi(x^2)&=&\Big(1+\frac{h}{2}\Big)(-h^2_{\phantom{.}\hat{0}}/2).
\ee

We write the partition function as $Z_{\rm ex}=\int \mathcal{D}\psi^{*}\mathcal{D}\psi\exp\left(-\mathcal{S}\right)$ with imaginary-time boundary action~\cite{stoof2009ultracold} $\mathcal{S}
= \int^{\beta}_0d\tau\int dx\psi^{*}(x^2,\tau)({\partial_\tau}+\mathcal{H}_{\rm ex})\psi(x^2,\tau)$. 
Integrating over fermionic fields, the effective free energy functional of the gravitational field is obtained as $\mathcal{F}[\zeta,\chi]=-\beta^{-1}\ln Z_{\rm ex}=\beta^{-1}\mathcal{S}[\zeta,\chi]$. Up to a constant which is independent of $\zeta$ and $\chi$, the effective action can be expressed as $\mathcal{S}[\zeta,\chi]=\sum_{l=1}^{\infty}\text{tr}[(G_0\Sigma)^l]/l$, where the trace is to be taken over the real space direction $x^2$ and over imaginary time $\tau$~\cite{supplemental_material}.
Moreover, since $\chi(x)$ equals zero for a gravitational potential of Luttinger's type, in the free energy we only need to consider the zeroth- and first-order term in powers of the variable $\chi$~\footnote{The variable $\chi$ should not be confused with the value $\chi(x)$ that it is substituted into the expression for the energy current {\it after} the functional derivative of the free energy has been taken.}, as higher-order terms will not contribute to Eq.~(\ref{eq:energy_current_expect}).

After some lengthy algebra~\cite{supplemental_material}, we obtain the following expression for the boundary free energy in the long-wavelength limit~\footnote{Since we neglected temperature-independent terms in our derivation, this form of the boundary free energy is only valid up to a constant which is independent of temperature.}
\be \label{eq:free_energy_result}
\mathcal{F}[\zeta,\chi]&=&\frac{\pi T^2}{12}\int dx \frac{\zeta(x)}{1+\zeta(x)}-\frac{7\pi^3 T^4}{60}\int dx \frac{\chi(x)}{(1+\zeta(x))^4}.
\nn
\ee
This equation is the second main finding of our work.

{\it Edge energy current}---Substituting Eq.~(\ref{eq:free_energy_result}) into Eq.~(\ref{eq:energy_current_expect}),  the energy current flowing along the boundary is found to be (restore $m_N$ from here)
\be \label{eq:energy_bdry_current}
j_E^{\text{bdry}}(x^2)&=&-\frac{\pi T^2}{12}-\frac{7\pi^3 T^4}{60m_N^2}\frac{1}{(1+\phi(x))^2}.
\ee

We stress that the long-wavelength limit has been taken throughout this paper. This means that we have neglected terms in the boundary current that depend on derivatives of the metric. As shown below, such terms would correspond to contributions to the bulk thermal Hall current that are proportional to at least the second derivative of the gravitational potential. Thus, these terms do not contribute to the linear response thermal Hall conductivity $\kappa_{xy}$.

In fact, conservation of energy currents requires the presence of a bulk thermal Hall current to compensate for anomalies at the boundary. In other words, if a (1+1)-dimensional current $j^{\text{bdry}}_E$ flows along the boundary, then the bulk current $j^{1}_E$ crossing the boundary must satisfy the continuity equation~\cite{nakai2016finite} 
\be \label{eq:bulk_boundary}
j_E^1(x^1=+0)=-\partial_2j_E^{\rm bdry}(x^2).
\ee
Thus, the result of Eq.~(\ref{eq:energy_bdry_current}) indicates that, relative to the massive Dirac fermions model~\cite{liu2024no}, the introduction of the Newtonian mass term leads to an additional higher-order term in the boundary current, causing the system to exhibit a finite bulk thermal current in the linear regime. According to the continuity equation~(\ref{eq:bulk_boundary}), the bulk thermal current reads
\be
j_E^{1}(x^1=+0)=-\frac{7\pi^3 T^4}{30m_N^2}\partial_2\phi(x^2).
\ee
Using linear response theory and the Tolman–Ehrenfest relation~\cite{luttinger1964theory,qin2011energy}, the thermal conductivity contributed by this term is $\kappa^{\text{bulk}}_{xy}=7\pi^3 T^3/30m_N^2$. 

Finally, we can use Eq.~(\ref{eq:energy_bdry_current}) to calculate the system's thermal Hall conductivity based on the generalized Streda formula~\cite{nomura2012cross,nakai2016finite}. According to the definition of energy magnetization $M_E^z$ in terms of energy current~\cite{cooper1997thermoelectric}, the boundary energy current satisfies the relation $j^{\text{bdry}}_E=- M_E^z(x^1=+\infty)$~\cite{nakai2016finite,zhang2020thermodynamics,supplemental_material}. From this, we get $M^z_E(\phi=0)=\pi T^2/12+7\pi^3 T^4/(60m_N^2)$. Therefore, the thermal Hall conductivity is given by~\cite{qin2011energy,guo2020gauge}
\be \label{eq:kappa_full}
\kappa_{xy}=\kappa_{xy}^{\text{bulk}}+\frac{2M_E}{T}=\frac{\pi T}{6}+\frac{7\pi^3 T^3}{15m_N^2}
.
\ee

This result is identical to that obtained directly using the generalized Streda formula~\cite{nomura2012cross,zhang2020thermodynamics}
\be
\kappa_{xy}=\frac{\partial M_E^z}{\partial T}=\frac{\pi T}{6}+\frac{7\pi^3T^3}{15m_N^2}
\ee

The first term in Eq.~(\ref{eq:kappa_full}) corresponds to a quantized thermal Hall conductivity with Chern number $C=-1$, and the second contribution, which breaks the Wiedemann-Franz law, corresponds to the bulk thermal current induced by the Newtonian mass term in the topological insulator's Hamiltonian. When the Newtonian mass term vanishes (as $m_N\rightarrow \infty$), the second contribution approaches zero.

{\it Conclusion}--- In this paper, we have developed a systematic approach based on an extended Hamiltonian to calculate the contribution of boundary states to the thermal Hall conductivity. Unlike previous approaches~\cite{stone2012gravitational,liu2024no,nakai2016finite,bradlyn2015low}, our method does not rely on the assumption of Lorentz symmetry. Although our derivation is based on the choice of a specific Hamiltonian, this method can be easily generalized to other models.

By systematically resumming all-order contributions in powers of the metric tensor in the long-wavelength limit, we have obtained a rigorous expression for the boundary free energy. From this, we derived the boundary current generated by a Luttinger-type gravitational potential. We found that, compared to the massive Dirac fermions model, the introduction of the Newtonian mass term significantly alters the quantum anomalous behavior of the boundary states, resulting in a finite bulk thermal current. The contribution of the latter breaks the Wiedemann-Franz law.

{\it Acknowledgments}--- The authors would like to thank Alexander E. Kazantsev for useful discussions. A.P. acknowledges support from the European Commission under the EU Horizon 2020 MSCA-RISE-2019 programme (project 873028 HYDROTRONICS) and from the Leverhulme Trust under the grant agreement RPG-2023-253. A.D.D.-I. acknowledges support from the Engineering and Physical Sciences Research Council, Grant No. EP/T517823/1.

\bibliography{E1}

\begin{thebibliography}{33}%
\makeatletter
\providecommand \@ifxundefined [1]{%
 \@ifx{#1\undefined}
}%
\providecommand \@ifnum [1]{%
 \ifnum #1\expandafter \@firstoftwo
 \else \expandafter \@secondoftwo
 \fi
}%
\providecommand \@ifx [1]{%
 \ifx #1\expandafter \@firstoftwo
 \else \expandafter \@secondoftwo
 \fi
}%
\providecommand \natexlab [1]{#1}%
\providecommand \enquote  [1]{``#1''}%
\providecommand \bibnamefont  [1]{#1}%
\providecommand \bibfnamefont [1]{#1}%
\providecommand \citenamefont [1]{#1}%
\providecommand \href@noop [0]{\@secondoftwo}%
\providecommand \href [0]{\begingroup \@sanitize@url \@href}%
\providecommand \@href[1]{\@@startlink{#1}\@@href}%
\providecommand \@@href[1]{\endgroup#1\@@endlink}%
\providecommand \@sanitize@url [0]{\catcode `\\12\catcode `\$12\catcode `\&12\catcode `\#12\catcode `\^12\catcode `\_12\catcode `\%12\relax}%
\providecommand \@@startlink[1]{}%
\providecommand \@@endlink[0]{}%
\providecommand \url  [0]{\begingroup\@sanitize@url \@url }%
\providecommand \@url [1]{\endgroup\@href {#1}{\urlprefix }}%
\providecommand \urlprefix  [0]{URL }%
\providecommand \Eprint [0]{\href }%
\providecommand \doibase [0]{https://doi.org/}%
\providecommand \selectlanguage [0]{\@gobble}%
\providecommand \bibinfo  [0]{\@secondoftwo}%
\providecommand \bibfield  [0]{\@secondoftwo}%
\providecommand \translation [1]{[#1]}%
\providecommand \BibitemOpen [0]{}%
\providecommand \bibitemStop [0]{}%
\providecommand \bibitemNoStop [0]{.\EOS\space}%
\providecommand \EOS [0]{\spacefactor3000\relax}%
\providecommand \BibitemShut  [1]{\csname bibitem#1\endcsname}%
\let\auto@bib@innerbib\@empty
\bibitem [{\citenamefont {Golan}\ and\ \citenamefont {Stern}(2018)}]{golan2018probing}%
  \BibitemOpen
  \bibfield  {author} {\bibinfo {author} {\bibfnamefont {O.}~\bibnamefont {Golan}}\ and\ \bibinfo {author} {\bibfnamefont {A.}~\bibnamefont {Stern}},\ }\bibfield  {title} {\bibinfo {title} {Probing topological superconductors with emergent gravity},\ }\href@noop {} {\bibfield  {journal} {\bibinfo  {journal} {Physical Review B}\ }\textbf {\bibinfo {volume} {98}},\ \bibinfo {pages} {064503} (\bibinfo {year} {2018})}\BibitemShut {NoStop}%
\bibitem [{\citenamefont {Grissonnanche}\ \emph {et~al.}(2020)\citenamefont {Grissonnanche}, \citenamefont {Th{\'e}riault}, \citenamefont {Gourgout}, \citenamefont {Boulanger}, \citenamefont {Lefran{\c{c}}ois}, \citenamefont {Ataei}, \citenamefont {Lalibert{\'e}}, \citenamefont {Dion}, \citenamefont {Zhou}, \citenamefont {Pyon} \emph {et~al.}}]{grissonnanche2020chiral}%
  \BibitemOpen
  \bibfield  {author} {\bibinfo {author} {\bibfnamefont {G.}~\bibnamefont {Grissonnanche}}, \bibinfo {author} {\bibfnamefont {S.}~\bibnamefont {Th{\'e}riault}}, \bibinfo {author} {\bibfnamefont {A.}~\bibnamefont {Gourgout}}, \bibinfo {author} {\bibfnamefont {M.-E.}\ \bibnamefont {Boulanger}}, \bibinfo {author} {\bibfnamefont {E.}~\bibnamefont {Lefran{\c{c}}ois}}, \bibinfo {author} {\bibfnamefont {A.}~\bibnamefont {Ataei}}, \bibinfo {author} {\bibfnamefont {F.}~\bibnamefont {Lalibert{\'e}}}, \bibinfo {author} {\bibfnamefont {M.}~\bibnamefont {Dion}}, \bibinfo {author} {\bibfnamefont {J.-S.}\ \bibnamefont {Zhou}}, \bibinfo {author} {\bibfnamefont {S.}~\bibnamefont {Pyon}}, \emph {et~al.},\ }\bibfield  {title} {\bibinfo {title} {Chiral phonons in the pseudogap phase of cuprates},\ }\href@noop {} {\bibfield  {journal} {\bibinfo  {journal} {Nature Physics}\ }\textbf {\bibinfo {volume} {16}},\ \bibinfo {pages} {1108} (\bibinfo {year} {2020})}\BibitemShut {NoStop}%
\bibitem [{\citenamefont {Kane}\ and\ \citenamefont {Fisher}(1997)}]{kane1997quantized}%
  \BibitemOpen
  \bibfield  {author} {\bibinfo {author} {\bibfnamefont {C.}~\bibnamefont {Kane}}\ and\ \bibinfo {author} {\bibfnamefont {M.~P.}\ \bibnamefont {Fisher}},\ }\bibfield  {title} {\bibinfo {title} {Quantized thermal transport in the fractional quantum hall effect},\ }\href@noop {} {\bibfield  {journal} {\bibinfo  {journal} {Physical Review B}\ }\textbf {\bibinfo {volume} {55}},\ \bibinfo {pages} {15832} (\bibinfo {year} {1997})}\BibitemShut {NoStop}%
\bibitem [{\citenamefont {Ideue}\ \emph {et~al.}(2017)\citenamefont {Ideue}, \citenamefont {Kurumaji}, \citenamefont {Ishiwata},\ and\ \citenamefont {Tokura}}]{ideue2017giant}%
  \BibitemOpen
  \bibfield  {author} {\bibinfo {author} {\bibfnamefont {T.}~\bibnamefont {Ideue}}, \bibinfo {author} {\bibfnamefont {T.}~\bibnamefont {Kurumaji}}, \bibinfo {author} {\bibfnamefont {S.}~\bibnamefont {Ishiwata}},\ and\ \bibinfo {author} {\bibfnamefont {Y.}~\bibnamefont {Tokura}},\ }\bibfield  {title} {\bibinfo {title} {Giant thermal hall effect in multiferroics},\ }\href@noop {} {\bibfield  {journal} {\bibinfo  {journal} {Nature materials}\ }\textbf {\bibinfo {volume} {16}},\ \bibinfo {pages} {797} (\bibinfo {year} {2017})}\BibitemShut {NoStop}%
\bibitem [{\citenamefont {Shimizu}\ \emph {et~al.}(2015)\citenamefont {Shimizu}, \citenamefont {Yamakage},\ and\ \citenamefont {Nomura}}]{shimizu2015quantum}%
  \BibitemOpen
  \bibfield  {author} {\bibinfo {author} {\bibfnamefont {Y.}~\bibnamefont {Shimizu}}, \bibinfo {author} {\bibfnamefont {A.}~\bibnamefont {Yamakage}},\ and\ \bibinfo {author} {\bibfnamefont {K.}~\bibnamefont {Nomura}},\ }\bibfield  {title} {\bibinfo {title} {Quantum thermal hall effect of majorana fermions on the surface of superconducting topological insulators},\ }\href@noop {} {\bibfield  {journal} {\bibinfo  {journal} {Physical Review B}\ }\textbf {\bibinfo {volume} {91}},\ \bibinfo {pages} {195139} (\bibinfo {year} {2015})}\BibitemShut {NoStop}%
\bibitem [{\citenamefont {Zhang}\ \emph {et~al.}(2024)\citenamefont {Zhang}, \citenamefont {Gao},\ and\ \citenamefont {Chen}}]{zhang2024thermal}%
  \BibitemOpen
  \bibfield  {author} {\bibinfo {author} {\bibfnamefont {X.-T.}\ \bibnamefont {Zhang}}, \bibinfo {author} {\bibfnamefont {Y.~H.}\ \bibnamefont {Gao}},\ and\ \bibinfo {author} {\bibfnamefont {G.}~\bibnamefont {Chen}},\ }\bibfield  {title} {\bibinfo {title} {Thermal hall effects in quantum magnets},\ }\href@noop {} {\bibfield  {journal} {\bibinfo  {journal} {Physics Reports}\ }\textbf {\bibinfo {volume} {1070}},\ \bibinfo {pages} {1} (\bibinfo {year} {2024})}\BibitemShut {NoStop}%
\bibitem [{\citenamefont {Grissonnanche}\ \emph {et~al.}(2019)\citenamefont {Grissonnanche}, \citenamefont {Legros}, \citenamefont {Badoux}, \citenamefont {Lefran{\c{c}}ois}, \citenamefont {Zatko}, \citenamefont {Lizaire}, \citenamefont {Lalibert{\'e}}, \citenamefont {Gourgout}, \citenamefont {Zhou}, \citenamefont {Pyon} \emph {et~al.}}]{grissonnanche2019giant}%
  \BibitemOpen
  \bibfield  {author} {\bibinfo {author} {\bibfnamefont {G.}~\bibnamefont {Grissonnanche}}, \bibinfo {author} {\bibfnamefont {A.}~\bibnamefont {Legros}}, \bibinfo {author} {\bibfnamefont {S.}~\bibnamefont {Badoux}}, \bibinfo {author} {\bibfnamefont {E.}~\bibnamefont {Lefran{\c{c}}ois}}, \bibinfo {author} {\bibfnamefont {V.}~\bibnamefont {Zatko}}, \bibinfo {author} {\bibfnamefont {M.}~\bibnamefont {Lizaire}}, \bibinfo {author} {\bibfnamefont {F.}~\bibnamefont {Lalibert{\'e}}}, \bibinfo {author} {\bibfnamefont {A.}~\bibnamefont {Gourgout}}, \bibinfo {author} {\bibfnamefont {J.-S.}\ \bibnamefont {Zhou}}, \bibinfo {author} {\bibfnamefont {S.}~\bibnamefont {Pyon}}, \emph {et~al.},\ }\bibfield  {title} {\bibinfo {title} {Giant thermal hall conductivity in the pseudogap phase of cuprate superconductors},\ }\href@noop {} {\bibfield  {journal} {\bibinfo  {journal} {Nature}\ }\textbf {\bibinfo {volume} {571}},\ \bibinfo {pages} {376} (\bibinfo {year} {2019})}\BibitemShut {NoStop}%
\bibitem [{\citenamefont {Samajdar}\ \emph {et~al.}(2019)\citenamefont {Samajdar}, \citenamefont {Scheurer}, \citenamefont {Chatterjee}, \citenamefont {Guo}, \citenamefont {Xu},\ and\ \citenamefont {Sachdev}}]{samajdar2019enhanced}%
  \BibitemOpen
  \bibfield  {author} {\bibinfo {author} {\bibfnamefont {R.}~\bibnamefont {Samajdar}}, \bibinfo {author} {\bibfnamefont {M.~S.}\ \bibnamefont {Scheurer}}, \bibinfo {author} {\bibfnamefont {S.}~\bibnamefont {Chatterjee}}, \bibinfo {author} {\bibfnamefont {H.}~\bibnamefont {Guo}}, \bibinfo {author} {\bibfnamefont {C.}~\bibnamefont {Xu}},\ and\ \bibinfo {author} {\bibfnamefont {S.}~\bibnamefont {Sachdev}},\ }\bibfield  {title} {\bibinfo {title} {Enhanced thermal hall effect in the square-lattice n{\'e}el state},\ }\href@noop {} {\bibfield  {journal} {\bibinfo  {journal} {Nature Physics}\ }\textbf {\bibinfo {volume} {15}},\ \bibinfo {pages} {1290} (\bibinfo {year} {2019})}\BibitemShut {NoStop}%
\bibitem [{\citenamefont {Fulga}\ \emph {et~al.}(2020)\citenamefont {Fulga}, \citenamefont {Oreg}, \citenamefont {Mirlin}, \citenamefont {Stern},\ and\ \citenamefont {Mross}}]{fulga2020temperature}%
  \BibitemOpen
  \bibfield  {author} {\bibinfo {author} {\bibfnamefont {I.}~\bibnamefont {Fulga}}, \bibinfo {author} {\bibfnamefont {Y.}~\bibnamefont {Oreg}}, \bibinfo {author} {\bibfnamefont {A.~D.}\ \bibnamefont {Mirlin}}, \bibinfo {author} {\bibfnamefont {A.}~\bibnamefont {Stern}},\ and\ \bibinfo {author} {\bibfnamefont {D.~F.}\ \bibnamefont {Mross}},\ }\bibfield  {title} {\bibinfo {title} {Temperature enhancement of thermal hall conductance quantization},\ }\href@noop {} {\bibfield  {journal} {\bibinfo  {journal} {Physical review letters}\ }\textbf {\bibinfo {volume} {125}},\ \bibinfo {pages} {236802} (\bibinfo {year} {2020})}\BibitemShut {NoStop}%
\bibitem [{\citenamefont {Cooper}\ \emph {et~al.}(1997)\citenamefont {Cooper}, \citenamefont {Halperin},\ and\ \citenamefont {Ruzin}}]{cooper1997thermoelectric}%
  \BibitemOpen
  \bibfield  {author} {\bibinfo {author} {\bibfnamefont {N.}~\bibnamefont {Cooper}}, \bibinfo {author} {\bibfnamefont {B.}~\bibnamefont {Halperin}},\ and\ \bibinfo {author} {\bibfnamefont {I.}~\bibnamefont {Ruzin}},\ }\bibfield  {title} {\bibinfo {title} {Thermoelectric response of an interacting two-dimensional electron gas in a quantizing magnetic field},\ }\href@noop {} {\bibfield  {journal} {\bibinfo  {journal} {Physical Review B}\ }\textbf {\bibinfo {volume} {55}},\ \bibinfo {pages} {2344} (\bibinfo {year} {1997})}\BibitemShut {NoStop}%
\bibitem [{\citenamefont {Qin}\ \emph {et~al.}(2011)\citenamefont {Qin}, \citenamefont {Niu},\ and\ \citenamefont {Shi}}]{qin2011energy}%
  \BibitemOpen
  \bibfield  {author} {\bibinfo {author} {\bibfnamefont {T.}~\bibnamefont {Qin}}, \bibinfo {author} {\bibfnamefont {Q.}~\bibnamefont {Niu}},\ and\ \bibinfo {author} {\bibfnamefont {J.}~\bibnamefont {Shi}},\ }\bibfield  {title} {\bibinfo {title} {Energy magnetization and the thermal hall effect},\ }\href@noop {} {\bibfield  {journal} {\bibinfo  {journal} {Physical review letters}\ }\textbf {\bibinfo {volume} {107}},\ \bibinfo {pages} {236601} (\bibinfo {year} {2011})}\BibitemShut {NoStop}%
\bibitem [{\citenamefont {Luttinger}(1964)}]{luttinger1964theory}%
  \BibitemOpen
  \bibfield  {author} {\bibinfo {author} {\bibfnamefont {J.}~\bibnamefont {Luttinger}},\ }\bibfield  {title} {\bibinfo {title} {Theory of thermal transport coefficients},\ }\href@noop {} {\bibfield  {journal} {\bibinfo  {journal} {Physical Review}\ }\textbf {\bibinfo {volume} {135}},\ \bibinfo {pages} {A1505} (\bibinfo {year} {1964})}\BibitemShut {NoStop}%
\bibitem [{\citenamefont {Shitade}(2014)}]{shitade2014heat}%
  \BibitemOpen
  \bibfield  {author} {\bibinfo {author} {\bibfnamefont {A.}~\bibnamefont {Shitade}},\ }\bibfield  {title} {\bibinfo {title} {Heat transport as torsional responses and keldysh formalism in a curved spacetime},\ }\href@noop {} {\bibfield  {journal} {\bibinfo  {journal} {Progress of Theoretical and Experimental Physics}\ }\textbf {\bibinfo {volume} {2014}},\ \bibinfo {pages} {123I01} (\bibinfo {year} {2014})}\BibitemShut {NoStop}%
\bibitem [{\citenamefont {Paul}\ and\ \citenamefont {Kotliar}(2003)}]{paul2003thermal}%
  \BibitemOpen
  \bibfield  {author} {\bibinfo {author} {\bibfnamefont {I.}~\bibnamefont {Paul}}\ and\ \bibinfo {author} {\bibfnamefont {G.}~\bibnamefont {Kotliar}},\ }\bibfield  {title} {\bibinfo {title} {Thermal transport for many-body tight-binding models},\ }\href@noop {} {\bibfield  {journal} {\bibinfo  {journal} {Physical review B}\ }\textbf {\bibinfo {volume} {67}},\ \bibinfo {pages} {115131} (\bibinfo {year} {2003})}\BibitemShut {NoStop}%
\bibitem [{\citenamefont {Cappelli}\ \emph {et~al.}(2002)\citenamefont {Cappelli}, \citenamefont {Huerta},\ and\ \citenamefont {Zemba}}]{cappelli2002thermal}%
  \BibitemOpen
  \bibfield  {author} {\bibinfo {author} {\bibfnamefont {A.}~\bibnamefont {Cappelli}}, \bibinfo {author} {\bibfnamefont {M.}~\bibnamefont {Huerta}},\ and\ \bibinfo {author} {\bibfnamefont {G.~R.}\ \bibnamefont {Zemba}},\ }\bibfield  {title} {\bibinfo {title} {Thermal transport in chiral conformal theories and hierarchical quantum hall states},\ }\href@noop {} {\bibfield  {journal} {\bibinfo  {journal} {Nuclear Physics B}\ }\textbf {\bibinfo {volume} {636}},\ \bibinfo {pages} {568} (\bibinfo {year} {2002})}\BibitemShut {NoStop}%
\bibitem [{\citenamefont {Bradlyn}\ and\ \citenamefont {Read}(2015)}]{bradlyn2015low}%
  \BibitemOpen
  \bibfield  {author} {\bibinfo {author} {\bibfnamefont {B.}~\bibnamefont {Bradlyn}}\ and\ \bibinfo {author} {\bibfnamefont {N.}~\bibnamefont {Read}},\ }\bibfield  {title} {\bibinfo {title} {Low-energy effective theory in the bulk for transport in a topological phase},\ }\href@noop {} {\bibfield  {journal} {\bibinfo  {journal} {Physical Review B}\ }\textbf {\bibinfo {volume} {91}},\ \bibinfo {pages} {125303} (\bibinfo {year} {2015})}\BibitemShut {NoStop}%
\bibitem [{\citenamefont {Stone}(2012)}]{stone2012gravitational}%
  \BibitemOpen
  \bibfield  {author} {\bibinfo {author} {\bibfnamefont {M.}~\bibnamefont {Stone}},\ }\bibfield  {title} {\bibinfo {title} {Gravitational anomalies and thermal hall effect in topological insulators},\ }\href@noop {} {\bibfield  {journal} {\bibinfo  {journal} {Physical Review B—Condensed Matter and Materials Physics}\ }\textbf {\bibinfo {volume} {85}},\ \bibinfo {pages} {184503} (\bibinfo {year} {2012})}\BibitemShut {NoStop}%
\bibitem [{\citenamefont {Nomura}\ \emph {et~al.}(2012)\citenamefont {Nomura}, \citenamefont {Ryu}, \citenamefont {Furusaki},\ and\ \citenamefont {Nagaosa}}]{nomura2012cross}%
  \BibitemOpen
  \bibfield  {author} {\bibinfo {author} {\bibfnamefont {K.}~\bibnamefont {Nomura}}, \bibinfo {author} {\bibfnamefont {S.}~\bibnamefont {Ryu}}, \bibinfo {author} {\bibfnamefont {A.}~\bibnamefont {Furusaki}},\ and\ \bibinfo {author} {\bibfnamefont {N.}~\bibnamefont {Nagaosa}},\ }\bibfield  {title} {\bibinfo {title} {Cross-correlated responses of topological superconductors and superfluids},\ }\href@noop {} {\bibfield  {journal} {\bibinfo  {journal} {Physical review letters}\ }\textbf {\bibinfo {volume} {108}},\ \bibinfo {pages} {026802} (\bibinfo {year} {2012})}\BibitemShut {NoStop}%
\bibitem [{\citenamefont {Read}\ and\ \citenamefont {Green}(2000)}]{read2000paired}%
  \BibitemOpen
  \bibfield  {author} {\bibinfo {author} {\bibfnamefont {N.}~\bibnamefont {Read}}\ and\ \bibinfo {author} {\bibfnamefont {D.}~\bibnamefont {Green}},\ }\bibfield  {title} {\bibinfo {title} {Paired states of fermions in two dimensions with breaking of parity and time-reversal symmetries and the fractional quantum hall effect},\ }\href@noop {} {\bibfield  {journal} {\bibinfo  {journal} {Physical Review B}\ }\textbf {\bibinfo {volume} {61}},\ \bibinfo {pages} {10267} (\bibinfo {year} {2000})}\BibitemShut {NoStop}%
\bibitem [{\citenamefont {Sumiyoshi}\ and\ \citenamefont {Fujimoto}(2013)}]{sumiyoshi2013quantum}%
  \BibitemOpen
  \bibfield  {author} {\bibinfo {author} {\bibfnamefont {H.}~\bibnamefont {Sumiyoshi}}\ and\ \bibinfo {author} {\bibfnamefont {S.}~\bibnamefont {Fujimoto}},\ }\bibfield  {title} {\bibinfo {title} {Quantum thermal hall effect in a time-reversal-symmetry-broken topological superconductor in two dimensions: approach from bulk calculations},\ }\href@noop {} {\bibfield  {journal} {\bibinfo  {journal} {Journal of the Physical Society of Japan}\ }\textbf {\bibinfo {volume} {82}},\ \bibinfo {pages} {023602} (\bibinfo {year} {2013})}\BibitemShut {NoStop}%
\bibitem [{\citenamefont {Nakai}\ \emph {et~al.}(2016)\citenamefont {Nakai}, \citenamefont {Ryu},\ and\ \citenamefont {Nomura}}]{nakai2016finite}%
  \BibitemOpen
  \bibfield  {author} {\bibinfo {author} {\bibfnamefont {R.}~\bibnamefont {Nakai}}, \bibinfo {author} {\bibfnamefont {S.}~\bibnamefont {Ryu}},\ and\ \bibinfo {author} {\bibfnamefont {K.}~\bibnamefont {Nomura}},\ }\bibfield  {title} {\bibinfo {title} {Finite-temperature effective boundary theory of the quantized thermal hall effect},\ }\href@noop {} {\bibfield  {journal} {\bibinfo  {journal} {New Journal of Physics}\ }\textbf {\bibinfo {volume} {18}},\ \bibinfo {pages} {023038} (\bibinfo {year} {2016})}\BibitemShut {NoStop}%
\bibitem [{\citenamefont {Liu}\ \emph {et~al.}(2024)\citenamefont {Liu}, \citenamefont {Dumitriu-I},\ and\ \citenamefont {Principi}}]{liu2024no}%
  \BibitemOpen
  \bibfield  {author} {\bibinfo {author} {\bibfnamefont {F.}~\bibnamefont {Liu}}, \bibinfo {author} {\bibfnamefont {A.~D.}\ \bibnamefont {Dumitriu-I}},\ and\ \bibinfo {author} {\bibfnamefont {A.}~\bibnamefont {Principi}},\ }\bibfield  {title} {\bibinfo {title} {No bulk thermal currents in massive dirac fermions},\ }\href@noop {} {\bibfield  {journal} {\bibinfo  {journal} {Physical Review B}\ }\textbf {\bibinfo {volume} {110}},\ \bibinfo {pages} {L081404} (\bibinfo {year} {2024})}\BibitemShut {NoStop}%
\bibitem [{\citenamefont {Qi}\ and\ \citenamefont {Zhang}(2011)}]{qi2011topological}%
  \BibitemOpen
  \bibfield  {author} {\bibinfo {author} {\bibfnamefont {X.-L.}\ \bibnamefont {Qi}}\ and\ \bibinfo {author} {\bibfnamefont {S.-C.}\ \bibnamefont {Zhang}},\ }\bibfield  {title} {\bibinfo {title} {Topological insulators and superconductors},\ }\href@noop {} {\bibfield  {journal} {\bibinfo  {journal} {Reviews of modern physics}\ }\textbf {\bibinfo {volume} {83}},\ \bibinfo {pages} {1057} (\bibinfo {year} {2011})}\BibitemShut {NoStop}%
\bibitem [{\citenamefont {Czajka}(2022)}]{czajka2022exotic}%
  \BibitemOpen
  \bibfield  {author} {\bibinfo {author} {\bibfnamefont {P.}~\bibnamefont {Czajka}},\ }\emph {\bibinfo {title} {Exotic Thermal Transport in a Kitaev Magnet}},\ \href@noop {} {Ph.D. thesis},\ \bibinfo  {school} {Princeton University} (\bibinfo {year} {2022})\BibitemShut {NoStop}%
\bibitem [{\citenamefont {Vinkler-Aviv}\ and\ \citenamefont {Rosch}(2018)}]{vinkler2018approximately}%
  \BibitemOpen
  \bibfield  {author} {\bibinfo {author} {\bibfnamefont {Y.}~\bibnamefont {Vinkler-Aviv}}\ and\ \bibinfo {author} {\bibfnamefont {A.}~\bibnamefont {Rosch}},\ }\bibfield  {title} {\bibinfo {title} {Approximately quantized thermal hall effect of chiral liquids coupled to phonons},\ }\href@noop {} {\bibfield  {journal} {\bibinfo  {journal} {Physical Review X}\ }\textbf {\bibinfo {volume} {8}},\ \bibinfo {pages} {031032} (\bibinfo {year} {2018})}\BibitemShut {NoStop}%
\bibitem [{\citenamefont {Gromov}\ and\ \citenamefont {Abanov}(2015)}]{gromov2015thermal}%
  \BibitemOpen
  \bibfield  {author} {\bibinfo {author} {\bibfnamefont {A.}~\bibnamefont {Gromov}}\ and\ \bibinfo {author} {\bibfnamefont {A.~G.}\ \bibnamefont {Abanov}},\ }\bibfield  {title} {\bibinfo {title} {Thermal hall effect and geometry with torsion},\ }\href@noop {} {\bibfield  {journal} {\bibinfo  {journal} {Physical review letters}\ }\textbf {\bibinfo {volume} {114}},\ \bibinfo {pages} {016802} (\bibinfo {year} {2015})}\BibitemShut {NoStop}%
\bibitem [{\citenamefont {Carroll}(2019)}]{carroll2019spacetime}%
  \BibitemOpen
  \bibfield  {author} {\bibinfo {author} {\bibfnamefont {S.~M.}\ \bibnamefont {Carroll}},\ }\href@noop {} {\emph {\bibinfo {title} {Spacetime and geometry}}}\ (\bibinfo  {publisher} {Cambridge University Press},\ \bibinfo {year} {2019})\BibitemShut {NoStop}%
\bibitem [{sup()}]{supplemental_material}%
  \BibitemOpen
  \href@noop {} {\bibinfo {title} {See supplemental material for details of the derivations.}}\BibitemShut {Stop}%
\bibitem [{\citenamefont {Stoof}\ \emph {et~al.}(2009)\citenamefont {Stoof}, \citenamefont {Gubbels},\ and\ \citenamefont {Dickerscheid}}]{stoof2009ultracold}%
  \BibitemOpen
  \bibfield  {author} {\bibinfo {author} {\bibfnamefont {H.~T.}\ \bibnamefont {Stoof}}, \bibinfo {author} {\bibfnamefont {K.~B.}\ \bibnamefont {Gubbels}},\ and\ \bibinfo {author} {\bibfnamefont {D.}~\bibnamefont {Dickerscheid}},\ }\href@noop {} {\emph {\bibinfo {title} {Ultracold quantum fields}}}\ (\bibinfo  {publisher} {Springer},\ \bibinfo {year} {2009})\BibitemShut {NoStop}%
\bibitem [{Note1()}]{Note1}%
  \BibitemOpen
  \bibinfo {note} {The variable $\chi $ should not be confused with the value $\chi (x)$ that it is substituted into the expression for the energy current {\protect \it after} the functional derivative of the free energy has been taken.}\BibitemShut {Stop}%
\bibitem [{Note2()}]{Note2}%
  \BibitemOpen
  \bibinfo {note} {Since we neglected temperature-independent terms in our derivation, this form of the boundary free energy is only valid up to a constant which is independent of temperature.}\BibitemShut {Stop}%
\bibitem [{\citenamefont {Zhang}\ \emph {et~al.}(2020)\citenamefont {Zhang}, \citenamefont {Gao},\ and\ \citenamefont {Xiao}}]{zhang2020thermodynamics}%
  \BibitemOpen
  \bibfield  {author} {\bibinfo {author} {\bibfnamefont {Y.}~\bibnamefont {Zhang}}, \bibinfo {author} {\bibfnamefont {Y.}~\bibnamefont {Gao}},\ and\ \bibinfo {author} {\bibfnamefont {D.}~\bibnamefont {Xiao}},\ }\bibfield  {title} {\bibinfo {title} {Thermodynamics of energy magnetization},\ }\href@noop {} {\bibfield  {journal} {\bibinfo  {journal} {Physical Review B}\ }\textbf {\bibinfo {volume} {102}},\ \bibinfo {pages} {235161} (\bibinfo {year} {2020})}\BibitemShut {NoStop}%
\bibitem [{\citenamefont {Guo}\ \emph {et~al.}(2020)\citenamefont {Guo}, \citenamefont {Samajdar}, \citenamefont {Scheurer},\ and\ \citenamefont {Sachdev}}]{guo2020gauge}%
  \BibitemOpen
  \bibfield  {author} {\bibinfo {author} {\bibfnamefont {H.}~\bibnamefont {Guo}}, \bibinfo {author} {\bibfnamefont {R.}~\bibnamefont {Samajdar}}, \bibinfo {author} {\bibfnamefont {M.~S.}\ \bibnamefont {Scheurer}},\ and\ \bibinfo {author} {\bibfnamefont {S.}~\bibnamefont {Sachdev}},\ }\bibfield  {title} {\bibinfo {title} {Gauge theories for the thermal hall effect},\ }\href@noop {} {\bibfield  {journal} {\bibinfo  {journal} {Physical Review B}\ }\textbf {\bibinfo {volume} {101}},\ \bibinfo {pages} {195126} (\bibinfo {year} {2020})}\BibitemShut {NoStop}%
\end{thebibliography}%

%

\onecolumngrid

\onecolumngrid

\newpage
\clearpage

\vspace{1cm}
\begin{center}
\textbf{\large Supplemental material: Broken Lorentz symmetry and violation of the Wiedemann-Franz law in topological insulators}
\end{center}

\setcounter{secnumdepth}{3}
\setlength\parindent{0pt}

\setcounter{equation}{0}
\setcounter{figure}{0}
\setcounter{table}{0}
\setcounter{page}{1}

\makeatletter
\renewcommand{\theequation}{S\arabic{equation}}
\renewcommand{\thefigure}{S\arabic{figure}}
\renewcommand{\thetable}{S\arabic{table}}
\renewcommand{\thesection}{S\arabic{section}}
\newcommand{\lpartial}{\overleftarrow{\partial}}
\newcommand{\rpartial}{\overrightarrow{\partial}}
\newcommand{\lnabla}{\overleftarrow{\nabla}}
\newcommand{\rnabla}{\overrightarrow{\nabla}}

\section{Energy current operator without Lorentz symmetry}
The model action of the (2+1)-dimensional topological insulator is given by
\be
S&=&\int dt d^2x\psi^{\dagger}\gamma^{0}\Big(\frac{i}{2}(\gamma^{\mu}\rpartial_{\mu}-\lpartial_{\mu}\gamma^{\mu})-(m-\eta^{ij}\lpartial_i\rpartial_j)
\Big)\psi=\int dt d^2x \psi^{\dagger}\frac{i}{2}(\rpartial_0-\lpartial_0)\psi-\int dt H
\nn
H&=&\int d^2 x \psi^{\dagger}\Big(-\frac{i}{2}\gamma^0\gamma^j\rpartial_j+\frac{i}{2}\gamma^0\lpartial_j\gamma^j+\gamma^0(m-\eta^{ij}\lpartial_i\rpartial_j)
\Big)\psi
\ee
with the Minkowski metric $\eta_{\alpha\beta}=\text{diag}(1,-1,-1)$ and $\eta_{ab}=\text{diag}(-1,-1)$. According to Ref.~\cite{bradlyn2015low} , the action with gravitational field is given by
\be \label{eq:d_Ham_gen}
S&=&\int dtd^2x \sqrt{g}\psi^{\dagger}\gamma^{\hat{0}}\Big(\frac{i}{2}(e^{\mu}_{*\alpha}\gamma^{\alpha}\rnabla_{\mu}-\lnabla_{\mu}\gamma^{\alpha}e^{\mu}_{*\alpha})-m+\eta^{a b}e^{\mu}_{*a}e^{\nu}_{*b}\lnabla_{\mu}\rnabla_{\nu}
\Big)\psi
\nn
&=&\int dt d^2x \sqrt{g}\psi^{\dagger}\gamma^{\hat{0}}\frac{i}{2}\Big(e^{0}_{*\alpha}\gamma^{\alpha}\rpartial_0-\lpartial_0\gamma^{\alpha}e^{0}_{*\alpha}
\Big)\psi-\int dt H,
\nn
H&=&\int d^2x \sqrt{g} \psi^{\dagger}\gamma^{\hat{0}}\Big( \frac{i}{2}e^{0}_{*\alpha}\gamma^{\alpha}\omega_0+\frac{i}{2}\omega_0\gamma^{\alpha}e^{0}_{*\alpha}-\frac{i}{2}e^{j}_{*\alpha}\gamma^{\alpha}\rnabla_j+\frac{i}{2}\lnabla_j\gamma^{\alpha}e^{j}_{*\alpha}+m-\eta^{ab}e^{\mu}_{*a}e^{\nu}_{*b}\lnabla_{\mu}\rnabla_{\nu}\Big)\psi.
\ee
The covariant derivative $\rnabla_{\mu}$ ($\lnabla_\mu$) acts on the right (left) two-component spinor field $\psi$,
\be
\rnabla_\mu \psi&=&\Big(\rpartial_\mu+\frac{1}{8}[\gamma^{\alpha},\gamma^{\beta}]\omega_{\mu\alpha\beta}
\Big)\psi\equiv (\rpartial_\mu-\omega_\mu)\psi
\nn
\bar{\psi}\lnabla_{\mu}&=&\bar{\psi}\Big(\lpartial_\mu-\frac{1}{8}[\gamma^{\alpha},\gamma^{\beta}]\omega_{\mu\alpha\beta}
\Big)\equiv\bar{\psi}(\lpartial_\mu+\omega_\mu)
\ee
Based on the definition given by Refs.~\cite{bradlyn2015low,gromov2015thermal}, the energy current in the Luttinger case is
\be \label{eq:d_energy_def}
j^i_E&=&\sqrt{g}\tau^{i}_{*\nu=0}=-e^{*\alpha}_0\frac{\delta S}{\delta e^{*\alpha}_i}=-e^{*\hat{0}}_{0}\frac{\delta S}{\delta e^{*\hat{0}}_i}
\nn
&=&\sqrt{g}\psi^{\dagger}\gamma^{\hat{0}}\Big(\frac{i}{2}\gamma^i\rnabla_0-\frac{i}{2}\lnabla_0\gamma^{i}+\eta^{ib}e^{\nu}_{*b}\lnabla_{0}\rnabla_{\nu}+\eta^{a i}e^{\mu}_{*a}\lnabla_\mu\rnabla_{0}
\Big)\psi
\ee
Here we use the fact that 
\be
\frac{\delta e^{0}_{*i}}{\delta e^{*\hat{0}}_{i}}=-\frac{1}{(1+\phi)}, \quad e^{0}_{*\hat{0}}=\frac{1}{1+\phi},\quad e^{*\hat{0}}_0=1+\phi,\quad \sqrt{g}=1+\phi
\ee

Since we have completed all the functional derivatives, we can represent the energy current in the Luttinger case (without Lorentz symmetry) as follows. 
\be \label{eq:d_energy_operator_action}
j^{i}_E&=&(1+\phi)\psi^{\dagger}\gamma^{\hat{0}}\Big(\frac{i}{2}\gamma^i\rpartial_0-\frac{i}{2}\lpartial_0\gamma^{i}+\eta^{ib}e^{\nu}_{*b}\lpartial_{0}\rpartial_{\nu}+\eta^{a i}e^{\mu}_{*a}\lpartial_\mu\rpartial_{0}
\Big)\psi
\nn
&=&(1+\phi)\psi^{\dagger}\gamma^{\hat{0}}\Big(\frac{i}{2}\gamma^i\rpartial_0-\frac{i}{2}\lpartial_0\gamma^{i}
\Big)\psi-(1+\phi)\psi^{\dagger}\gamma^{\hat{0}}\Big(\lpartial_{0}\rpartial_{i}+\lpartial_{i}\rpartial_{0}\Big)\psi
\ee
and the form of action and the Hamiltonian becomes (with definition $\phi_g=1+\phi$)
\be
S&=&\int dt d^2x \frac{1+\phi}{1+\phi} \psi^{\dagger}\frac{i}{2}(\rpartial_0-\lpartial_0)\psi-\int dt d^2x (1+\phi)  \psi^{\dagger}\gamma^{\hat{0}}\Big( -\frac{i}{2}\gamma^{j}\rpartial_j+\frac{i}{2}\lpartial_j\gamma^{j}+m-\eta^{ab}\lpartial_{a}\rpartial_{b}\Big)\psi
\nn
&=&\int dt \bar{\psi}\frac{i}{2}(\gamma^{0}\rpartial_0-\lpartial_0\gamma^{0})\psi-\int dtd^2 x \phi_g \bar{\psi}\Big( -\frac{i}{2}\gamma^{j}\rpartial_j+\frac{i}{2}\lpartial_j\gamma^{j}+m+\lpartial_{1}\rpartial_{1}+\lpartial_{2}\rpartial_{2}\Big)\psi
\nn
H&=&\int dtd^2 x \phi_g \psi^{\dagger}\gamma^{\hat{0}}\Big( -\frac{i}{2}\gamma^{j}\rpartial_j+\frac{i}{2}\lpartial_j\gamma^{j}+m+\lpartial_{1}\rpartial_{1}+\lpartial_{2}\rpartial_{2}\Big)\psi.
\ee
The Heisenberg equation for $\psi$ (and $\psi^{\dagger}$) immediately yields the equation of motion, which is given by
\be \label{eq:d_eom}
i\partial_0\psi=[\psi,H]&=&\phi_g\gamma^{\hat{0}}\Big(-i\gamma^{j}\rpartial_j+m-\rpartial^2\Big)\psi-\frac{i}{2}\gamma^{\hat{0}}(\partial_j\phi_g)\gamma^{j}\psi-\gamma^{\hat{0}}\Big((\partial_1\phi_g)\rpartial_1+(\partial_2\phi_g)\rpartial_2\Big)\psi
\nn
i\partial_0\psi^{\dagger}=[\psi^{\dagger},H]&=&-\phi_g\psi^{\dagger}\gamma^{\hat{0}}\Big(i\lpartial_j\gamma^j+m-\lpartial^2\Big)-\frac{i}{2}\psi^{\dagger}\gamma^{\hat{0}}(\partial_j\phi_g)\gamma^j+\psi^{\dagger}\gamma^{\hat{0}}\Big(\lpartial_1(\partial_1\phi_g)+\lpartial_2(\partial_2\phi_g)\Big)
\ee
Here we use the identity $[A,BC]=\{A,B\}C-B\{A,C\}$ and define $\phi_g=1+\phi$. Substituting these results into the first part of Eq.~(\ref{eq:d_energy_operator_action}),we can express the first part of the energy current as follows:
\be
I&=&\frac{i}{2}\phi_g\psi^{\dagger}\gamma^{\hat{0}}\gamma^i\rpartial_0\psi
\nn
&=&\frac{1}{2}\phi_g^2\psi^{\dagger}\gamma^{\hat{0}}\gamma^{i}\gamma^{\hat{0}}\Big(-i\gamma^{j}\rpartial_j+m-\rpartial^2\Big)\psi-\frac{1}{2}\phi_g(\partial_j\phi_g)\psi^{\dagger}\gamma^{\hat{0}}\gamma^i\frac{i}{2}\gamma^{\hat{0}}\gamma^{j}\psi
-\frac{1}{2}\phi_g\psi^{\dagger}\gamma^{\hat{0}}\gamma^i\gamma^{\hat{0}}\Big((\partial_1\phi_g)\rpartial_1+(\partial_2\phi_g)\rpartial_2\Big)\psi
\nn
&=&-\frac{1}{2}\phi_g\psi^{\dagger}\gamma^i\Big(-i\phi_g\gamma^j\rpartial_j+\phi_gm -\phi_g\rpartial^2-\frac{i}{2}(\partial_j\phi_g)\gamma^j-(\partial_j\phi_g)\rpartial_j
\Big)\psi
\nn
II&=&-\frac{i}{2}\phi_g\psi^{\dagger}\gamma^{\hat{0}}\lpartial_0\gamma^{i}\psi=\frac{1}{2}\phi_g\psi^{\dagger}\gamma^{\hat{0}}\Big(\phi_gi\lpartial_j\gamma^j+\phi_gm-\phi_g\lpartial^2+\frac{i}{2}(\partial_j\phi_g)\gamma^j-\lpartial_j(\partial_j\phi_g)
\Big)\gamma^{\hat{0}}\gamma^i\psi
\nn
&=&\frac{1}{2}\phi_g\psi^{\dagger}\Big(-\phi_gi\lpartial_j\gamma^j+\phi_gm-\phi_g\lpartial^2-\frac{i}{2}(\partial_j\phi_g)\gamma^j-\lpartial_j(\partial_j\phi_g)
\Big)\gamma^i\psi
\ee
and
\be
I+II&=&\frac{i}{2}\phi_g^2\psi^{\dagger}(\gamma^{i}\gamma^{j}\rpartial_j-\lpartial_j\gamma^j\gamma^i)\psi-\frac{1}{2}\phi_g^2\psi^{\dagger}(\gamma^i m-m\gamma^i)\psi+\frac{1}{2}\phi_g^2\psi^{\dagger}\gamma^i(\rpartial^2-\lpartial^2)\psi
\nn
&&+\frac{i}{4}\phi_g(\partial_j\phi_g)\psi^{\dagger}(\gamma^i\gamma^j-\gamma^j\gamma^i)\psi+\frac{1}{2}\phi_g(\partial_j\phi_g)\psi^{\dagger}(\gamma^i\rpartial_j-\lpartial_j\gamma^i)\psi
\nn
&=&-\frac{i}{2}\phi_g^2\psi^{\dagger}(\rpartial_i-\lpartial_i)\psi+\frac{i}{2}\phi_g^2\psi^{\dagger}\gamma^{i}\gamma^{\bar{i}}(\rpartial_{\bar{i}}+\lpartial_{\bar{i}})\psi+\frac{1}{2}\phi_g^2\partial_j\Big(\psi^{\dagger}\gamma^i(\rpartial_j-\lpartial_j)\psi
\Big)
\nn
&&+\frac{i}{2}\phi_g(\partial_{\bar{i}}\phi_g)\psi^{\dagger}\gamma^i\gamma^{\bar{i}}\psi+\frac{1}{2}\phi_g(\partial_j\phi_g)\psi^{\dagger}\gamma^i(\rpartial_j-\lpartial_j)\psi
\nn
&=&-\frac{i}{2}\phi_g^2\psi^{\dagger}(\rpartial_i-\lpartial_i)\psi+\frac{i}{2}\phi_g\partial_{\bar{i}}\Big(\phi_g\psi^{\dagger}\gamma^{i}\gamma^{\bar{i}}\psi
\Big)+\frac{1}{2}\phi_g\partial_j\Big(\phi_g\psi^{\dagger}\gamma^i(\rpartial_j-\lpartial_j)\psi
\Big)
\ee
For the second part in Eq.~(\ref{eq:d_energy_operator_action}), we will not explicitly write the equations of motion to avoid excessive complexity. However, the time derivatives in Eq.~(\ref{eq:d_energy_opertor_em}) should be understood as corresponding to the equations of motion Eq.~(\ref{eq:d_eom}).
\be \label{eq:d_energy_opertor_em}
j^E_i&=&-\frac{i}{2}\phi_g^2\psi^{\dagger}(\rpartial_i-\lpartial_i)\psi+\frac{i}{2}\phi_g\partial_{\bar{i}}\Big(\phi_g\psi^{\dagger}\gamma^{i}\gamma^{\bar{i}}\psi
\Big)+\frac{1}{2}\phi_g\partial_j\Big(\phi_g\psi^{\dagger}\gamma^i(\rpartial_j-\lpartial_j)\psi
\Big)-\phi_g\psi^{\dagger}\gamma^{\hat{0}}\Big(\lpartial_{0}\rpartial_{i}+\lpartial_{i}\rpartial_{0}\Big)\psi
\nn
&=&-\frac{i}{2}\phi_g^2\psi^{\dagger}(\rpartial_i-\lpartial_i)\psi+\frac{1}{2}\phi_g\partial_{\bar{i}}\Big(\phi_g\psi^{\dagger}\Big(i\gamma^{i}\gamma^{\bar{i}}+\gamma^i(\rpartial_{\bar{i}}-\lpartial_{\bar{i}})\Big)\psi
\Big)+\frac{1}{2}\phi_g\partial_i\Big(\phi_g\psi^{\dagger}\gamma^i(\rpartial_i-\lpartial_i)\psi
\Big)
\nn
&&-\phi_g\psi^{\dagger}\gamma^{\hat{0}}\Big(\lpartial_{0}\rpartial_{i}+\lpartial_{i}\rpartial_{0}\Big)\psi
\ee

For systems with Lorentz invariance, due to the symmetry of the energy-momentum tensor, the expression of the energy current in Eq.~(\ref{eq:d_energy_opertor_em}) can equivalently be expressed as the functional derivative of the Hamiltonian~(\ref{eq:d_Ham_gen}). However, for the topological insulator system under consideration, this equivalence breaks down due to the absence of Lorentz symmetry. Nevertheless, as we demonstrate below, this equivalence can still be preserved if the Hamiltonian is appropriately extended.

To illustrate this point, we first consider the second and third terms in Eq.~(\ref{eq:d_energy_opertor_em}). These two terms can, in fact, be equivalently expressed as the spin current. We now define the time-independent spin current as follows:
\be
J^{\mu a b}_s&=&\frac{1}{\sqrt{g}}\frac{\delta S}{\delta\omega_{\mu a b}}=-\frac{1}{\sqrt{g}}\frac{\delta}{\delta \omega_{\mu a b}}\int dt H=-\frac{1}{\sqrt{g}}\frac{\delta H}{\delta \omega_{\mu a b}}
\nn
J^{i\hat{0}a}_s&=&-\psi^{\dagger}\gamma^{\hat{0}}\Big(-\frac{i}{2}\gamma^{i}\frac{1}{2}\gamma^{\hat{0}}\gamma^{a}+\frac{i}{2}\frac{-1}{2}\gamma^{\hat{0}}\gamma^{a}\gamma^{i}-\frac{1}{2}\gamma^{\hat{0}}\gamma^a\rnabla_i+\lnabla_i\frac{1}{2}\gamma^{\hat{0}}\gamma^a
\Big)\psi
\nn
&=&\psi^{\dagger}\Big(-\frac{i}{4}\gamma^{i}\gamma^{a}+\frac{i}{4}\gamma^a\gamma^{i}+\frac{1}{2}\gamma^a\rpartial_i-\frac{1}{2}\lpartial_i\gamma^{a}
\Big)\psi
\ee
As a result, we can express the energy current as
\be \label{eq:d_spin_cuurent}
&&\frac{1}{2}\phi_g\partial_{\bar{i}}\Big(\phi_g\psi^{\dagger}\Big(i\gamma^{i}\gamma^{\bar{i}}+\gamma^i(\rpartial_{\bar{i}}-\lpartial_{\bar{i}})\Big)\psi
\Big)=\phi_g\partial_{\bar{i}}\Big(\phi_gJ_s^{\bar{i}\hat{0}i}\Big)=-\phi_g\partial_{\bar{i}}\Big(\frac{\delta H}{\delta \omega_{\bar{i}\hat{0}i}}
\Big)
\nn
&&\frac{1}{2}\phi_g\partial_i\Big(\phi_g\psi^{\dagger}\gamma^i(\rpartial_i-\lpartial_i)\psi\Big)=\phi_g\partial_i\Big(\phi_gJ^{i\hat{0}i}
\Big)=-\phi_g\partial_i\Big(\frac{\delta H}{\delta \omega_{i\hat{0}i}}\Big)
\ee
Note that in Eq.~(\ref{eq:d_spin_cuurent}), identical indices do not imply summation. In this way, we have successfully expressed the second and third terms in Eq.~(\ref{eq:d_energy_opertor_em}) in terms of the spin current. Now, consider the last term in Eq.~(\ref{eq:d_energy_opertor_em}). From the derivation process of this, it can be observed that this term essentially originates from the explicit breaking of Lorentz symmetry in the quadratic terms of the action, namely:
\be \label{ed:d_Ham_ex}
\psi^{\dagger}\gamma^{\hat{0}}\Big(-\eta^{ab}e^{\mu}_{*a}e^{\nu}_{*b}\lnabla_{\mu}\rnabla_{\nu}\Big)\psi
&=&-\psi^{\dagger}\gamma^{\hat{0}}\Big(\eta^{\alpha\beta}e^{\mu}_{*\alpha}e^{\nu}_{*\beta}-\eta^{\hat{0}\hat{0}}e^{\mu}_{*\hat{0}}e^{\nu}_{*\hat{0}}-
\eta^{a\hat{0}}e^{\mu}_{*a}e^{\nu}_{*\hat{0}}-\eta^{\hat{0}b}e^{\mu}_{*\hat{0}}e^{\nu}_{*b}
\Big)\lnabla_{\mu}\rnabla_{\nu}\psi
\nn
&=&-\psi^{\dagger}\gamma^{\hat{0}}\Big(\eta^{\alpha\beta}e^{\mu}_{*\alpha}e^{\nu}_{*\beta}-\eta^{\hat{0}\hat{0}}e^{\mu}_{*\hat{0}}e^{\nu}_{*\hat{0}}
\Big)\lnabla_{\mu}\rnabla_{\nu}\psi
\nn
&=&-\psi^{\dagger}\gamma^{\hat{0}}\Big(\eta^{\alpha\beta}e^{\mu}_{*\alpha}e^{\nu}_{*\beta}
\Big)\lnabla_{\mu}\rnabla_{\nu}\psi-\psi^{\dagger}\gamma^{\hat{0}}\Big(-\eta^{\hat{0}\hat{0}}e^{0}_{*\hat{0}}e^{0}_{*\hat{0}}\lnabla_0\rnabla_0\Big)\psi
\nn
&&-\psi^{\dagger}\gamma^{\hat{0}}\Big(-\eta^{\hat{0}\hat{0}}e^{i}_{*\hat{0}}e^{0}_{*\hat{0}}\lnabla_i\rnabla_0-\eta^{\hat{0}\hat{0}}e^{0}_{*\hat{0}}e^{i}_{*\hat{0}}\lnabla_0\rnabla_i\Big)\psi
\ee
We replace the last term in Eq.~(\ref{eq:d_energy_opertor_em}) with the first term on the right-hand side of Eq.~(\ref{ed:d_Ham_ex}), Lorentz symmetry will be restored, and the expression of the energy current in Eq.~(\ref{eq:d_energy_opertor_em}) can equivalently be expressed as the functional derivative of Hamiltonian.
Now, the main task is to analyze the additional second, third, and fourth terms on the right-hand side of Eq.~(\ref{ed:d_Ham_ex}). From the definition of the energy-momentum tensor in Eq.~(\ref{eq:d_energy_def}), it is not difficult to see that, in the Luttinger case, the functional derivative corresponding to the second term in Eq.~(\ref{ed:d_Ham_ex}) equals zero; 
\be
\frac{\delta}{\delta e^{*\hat{0}}_i}\Big(\psi^{\dagger}\gamma^{\hat{0}}\eta^{\hat{0}\hat{0}}e^{0}_{*\hat{0}}e^{0}_{*\hat{0}}\lnabla_{0}\rnabla_{0}\psi\Big)=0=\frac{\delta}{\delta e^{*i}_{0}}\Big(\psi^{\dagger}\gamma^{\hat{0}}\eta^{\hat{0}\hat{0}}e^{0}_{*\hat{0}}e^{0}_{*\hat{0}}\lnabla_{0}\rnabla_{0}\psi\Big)
\ee
while the third and fourth terms in Eq.~(\ref{ed:d_Ham_ex}) are also zero in the Luttinger case. Based on this, we can construct these two terms as source terms in the extended Hamiltonian $H_{\rm ex}=H+H_{\rm src}$, which will generate the correct expression of the energy current.
\be
H_{\rm src}&=&\int d^2x \sqrt{g}\psi^{\dagger}\gamma^{\hat{0}}\Big(-\eta^{\hat{0}\hat{0}}e^{i}_{*\hat{0}}e^{0}_{*\hat{0}}\lpartial_i\rpartial_0-\eta^{\hat{0}\hat{0}}e^{0}_{*\hat{0}}e^{i}_{*\hat{0}}\lpartial_0\rpartial_i\Big)\psi
\ee
It is worth noting that the additional terms in this Hamiltonian do not affect the equations of motion, as these terms are zero in Luttinger gravity. This means that the extra terms belong to the external sources in the generating functional. Furthermore, the time derivatives in the extra terms are not actual derivatives but correspond to expressions of the equations of motion, similar with Eq.~(\ref{eq:d_energy_opertor_em}).

On the basis of this information, we can express the energy current as the functional derivative of the extended Hamiltonian. To demonstrate this, we directly calculate the energy current corresponding to the dual-energy-momentum tensor $\tau^{0}_{*i}$ in this case.
\be \label{eq:d_sec_energy}
\sqrt{g}\tau^{0}_{*i}&=&\sqrt{g}e^{*\alpha}_i\tau^{0}_{*\alpha}=\sqrt{g}\tau^{0}_{*i}=-\frac{\delta S_{\rm ex}}{\delta e^{*i}_0}=\frac{\delta}{\delta e^{*i}_0}\int dt H_{\rm ex}=\frac{\delta H_{\rm ex}}{\delta e^{*i}_0}
\nn
&=&\sqrt{g}\psi^{\dagger}\gamma^{\hat{0}}\Big(\frac{i}{2\phi_g}\gamma^{\hat{0}}\rnabla_i-\frac{i}{2\phi_g}\lnabla_i\gamma^{\hat{0}}+\frac{1}{\phi_g}e^{0}_{*\hat{0}}\lpartial_i\rpartial_0+\frac{1}{\phi_g}e^{0}_{*\hat{0}}\lpartial_0\rpartial_i
\Big)\psi
\nn
&=&\psi^{\dagger}\Big(\frac{i}{2}\rpartial_i-\frac{i}{2}\lpartial_i+\phi_g^{-1}\gamma^{\hat{0}}\lpartial_i\rpartial_0+\phi_g^{-1}\gamma^{\hat{0}}\lpartial_0\rpartial_i
\Big)\psi
\ee
Here, we used the relations,
\be
\frac{\delta e^{\nu}_{*\beta}}{\delta e^{*\alpha}_{\mu}}=-e^{\mu}_{*\beta}e^{\nu}_{*\alpha},\quad \frac{\delta e^{0}_{*\alpha}}{\delta e^{*i}_{0}}=0,\quad \frac{\delta }{\delta e^{*i}_{0}}\sqrt{g}=0
\ee
Combining Eqs.~(\ref{eq:d_energy_opertor_em}), (\ref{eq:d_spin_cuurent}), and (\ref{eq:d_sec_energy}), we can equivalently express the energy current as:
\be
j^E_i&=&-\phi_g^2\frac{\delta H_{\rm ex}}{\delta e^{*i}_0}-\phi_g\partial_{\bar{i}}\Big(\frac{\delta H}{\delta \omega_{\bar{i}\hat{0}i}}\Big)
-\phi_g\partial_i\Big(\frac{\delta H}{\delta \omega_{i\hat{0}i}}\Big)
=-\phi_g^2\frac{\delta H_{\rm ex}}{\delta e^{*i}_0}-\phi_g\partial_{\bar{i}}\Big(\frac{\delta H_{\rm ex}}{\delta \omega_{\bar{i}\hat{0}i}}\Big)
-\phi_g\partial_i\Big(\frac{\delta H_{\rm ex}}{\delta \omega_{i\hat{0}i}}\Big)
\ee
This relationship holds at the operator level, and identical indices do not imply summation.
\be \label{eq:d_enegy_curr_deriv}
j^E_1&=&-\phi_g^2\frac{\delta H_{\rm ex}}{\delta e^{*\hat{1}}_0}-\phi_g\partial_{2}\Big(\frac{\delta H_{\rm ex}}{\delta \omega_{2\hat{0}\hat{1}}}\Big)
-\phi_g\partial_1\Big(\frac{\delta H_{\rm ex}}{\delta \omega_{1\hat{0}\hat{1}}}\Big)
\nn
j^E_2&=&-\phi_g^2\frac{\delta H_{\rm ex}}{\delta e^{*\hat{2}}_0}-\phi_g\partial_{1}\Big(\frac{\delta H_{\rm ex}}{\delta \omega_{1\hat{0}\hat{2}}}\Big)
-\phi_g\partial_2\Big(\frac{\delta H_{\rm ex}}{\delta \omega_{2\hat{0}\hat{2}}}\Big)
\ee
Thus, at a finite temperature $1/\beta$, the expectation value of the energy current $j^E_2$ is given by
\be  \label{eq:d_enegy_curr_free_deriv}
j_E^2(x)&=&-\phi_g^2(x)\frac{1}{Z}\text{Tr}\Big[e^{-\beta H}  \frac{\delta H_{\rm ex}}{\delta e^{*\hat{2}}_0(x)}
\Big]-\sum_{i=1,2}\phi_g(x)\partial_i\Big(\frac{1}{Z}\text{Tr}\Big[e^{-\beta H}\frac{\delta H_{\rm ex}}{\delta \omega_{i\hat{0}\hat{2}}(x)}
\Big]\Big)
\nn
&=&-\phi_g^2(x)\frac{1}{Z_{\rm ex}}\text{Tr}\Big[e^{-\beta H_{\rm ex}}  \frac{\delta H_{\rm ex}}{\delta e^{*\hat{2}}_0(x)}
\Big]-\sum_{i=1,2}\phi_g(x)\partial_i\Big(\frac{1}{Z_{\rm ex}}\text{Tr}\Big[e^{-\beta H_{\rm ex}}\frac{\delta H_{\rm ex}}{\delta \omega_{i\hat{0}\hat{2}}(x)}
\Big]\Big)
\nn
&=&-\phi_g^2(x)\frac{\delta F_{\rm ex}}{\delta e^{*\hat{2}}_0(x)}-\sum_{i=1,2}\phi_g(x)\partial_i\Big(\frac{\delta F_{\rm ex}}{\delta \omega_{i\hat{0}\hat{2}}(x)}
\Big)
\ee
Here, we have introduced the extended partition function
\be \label{eq:d_part_function_s}
&&Z_{\rm ex} = \text{Tr}\left(e^{-\beta H_{\rm ex}}\right)=\int \mathcal{D}[\psi^*,\psi]\exp\left(-\int_0^{\beta}d\tau\int d^2x\psi^{*}(x,\tau)\frac{\partial}{\partial \tau}\psi(x,\tau)+H_{\rm ex}[\psi^*(\tau),\psi(\tau)]
\right)
\ee
and the free energy according to the usual relation $F_{\rm ex} = -\beta^{-1} \ln Z_{\rm ex}$. In the above derivation, we utilized the identity.
\be
\frac{1}{Z}\text{Tr}\left(e^{-\beta H}\frac{\delta H_{\rm ex}}{\delta e^{\phantom{.}\hat{2}}_{0}(x)}\Bigg|_{e^{\phantom{.}\hat{0}}_{\mu}=\delta^{\phantom{.}\hat{0}}_{\mu}\phi_g,
e^{\phantom{.}a}_{\mu}=\delta^{\phantom{.}a}_{\mu}}
\right)
=\frac{1}{Z_{\rm e}}\text{Tr}\left(e^{-\beta H_{\rm e}}\frac{\delta H_{\rm ex}}{\delta e^{\phantom{.}\hat{2}}_{0}(x)}
\right)\Bigg|_{e^{\phantom{.}\hat{0}}_{\mu}=\delta^{\phantom{.}\hat{0}}_{\mu}\phi_g,
e^{\phantom{.}a}_{\mu}=\delta^{\phantom{.}a}_{\mu}}.
\ee

\section{Boundary effective Hamiltonian}
In this section, we use the method provided in~\cite{nakai2016finite} to establish the effective theory for boundary states with the extended Hamiltonian. Since our ultimate goal is to use the functional derivative relationship given in the last section to derive the energy current $j_i^E$, we can first simplify the Hamiltonian by removing the parts that clearly do not contribute. In this case, the extended Hamiltonian $H_{\rm ex}=H+H_{\rm src}$ can be restated in the following form:
\be
H&=&\int d^2x \sqrt{g} \psi^{\dagger}\gamma^{\hat{0}}\Big( \frac{i}{2}e^{0}_{*\alpha}\gamma^{\alpha}\omega_0+\frac{i}{2}\omega_0\gamma^{\alpha}e^{0}_{*\alpha}-\frac{i}{2}e^{j}_{*\alpha}\gamma^{\alpha}\rnabla_j+\frac{i}{2}\lnabla_j\gamma^{\alpha}e^{j}_{*\alpha}+m-\eta^{ab}e^{\mu}_{*a}e^{\nu}_{*b}\lnabla_{\mu}\rnabla_{\nu}\Big)\psi
\nn
&=&\int d^2x \sqrt{g} \psi^{\dagger}\gamma^{\hat{0}}\Big( -\frac{i}{2}e^{j}_{*\alpha}\gamma^{\alpha}(\rpartial_j-\omega_j)+\frac{i}{2}(\lpartial_j+\omega_j)\gamma^{\alpha}e^{j}_{*\alpha}+m-\eta^{ab}e^{i}_{*a}e^{j}_{*b}(\lpartial_{i}+\omega_i)(\rpartial_{j}-\omega_j)\Big)\psi
\nn
H_{\rm src}&=&\int d^2x \sqrt{g}\psi^{\dagger}\gamma^{\hat{0}}\Big(-e^{i}_{*\hat{0}}e^{0}_{*\hat{0}}\lpartial_i\rpartial_0-e^{0}_{*\hat{0}}e^{i}_{*\hat{0}}\lpartial_0\rpartial_i\Big)\psi
\nn
&=&i\int d^2x \sqrt{g}e^{i}_{*\hat{0}}e^{0}_{*\hat{0}}\psi^{\dagger}\gamma^{\hat{0}}\lpartial_i\gamma^{\hat{0}}
\Big[\phi_g\Big(-i\gamma^{j}\rpartial_j+m-\rpartial^2\Big)-\frac{i}{2}(\partial_j\phi_g)\gamma^{j}-\Big((\partial_1\phi_g)\rpartial_1+(\partial_2\phi_g)\rpartial_2\Big)\Big]\psi
\nn
&&-i\int d^2x \sqrt{g}e^{0}_{*\hat{0}}e^{i}_{*\hat{0}}
\psi^{\dagger}\gamma^{\hat{0}}\Big[\phi_g\Big(i\lpartial_j\gamma^j+m-\lpartial^2\Big)+\frac{i}{2}(\partial_j\phi_g)\gamma^j-\Big(\lpartial_1(\partial_1\phi_g)+\lpartial_2(\partial_2\phi_g)\Big)\Big]\gamma^{\hat{0}}\rpartial_i\psi
\nn
&=&i\int d^2x e^{i}_{*\hat{0}}\psi^{\dagger}\lpartial_i
\Big[\phi_g\Big(-i\gamma^{j}\rpartial_j+m-\rpartial^2\Big)-\frac{i}{2}(\partial_j\phi_g)\gamma^{j}-(\partial_i\phi_g)\rpartial_i\Big]\psi
\nn
&&-i\int d^2x e^{i}_{*\hat{0}}
\psi^{\dagger}\Big[\phi_g\Big(-i\lpartial_j\gamma^j+m-\lpartial^2\Big)-\frac{i}{2}(\partial_j\phi_g)\gamma^j-\lpartial_i(\partial_i\phi_g)\Big]\rpartial_i\psi
\ee
Here, we used the relations,
\be \label{eq:d_deriv}
\frac{\delta e^{\nu}_{*\beta}}{\delta e^{*\alpha}_{\mu}}=-e^{\mu}_{*\beta}e^{\nu}_{*\alpha},\quad \frac{\delta e^{0}_{*\alpha}}{\delta e^{*i}_{0}}=0,\quad \frac{\delta }{\delta e^{*i}_{0}}\sqrt{g}=0,\quad
e^{0}_{*\hat{0}}=\frac{1}{1+\phi},\quad e^{*\hat{0}}_0=1+\phi,\quad \sqrt{g}=1+\phi.
\ee

Now we consider a boundary at $x^1=0$ between a topological insulator with mass $m<0$ (\textit{i.e.} in the half-plane $x^1>0$) and vacuum (\textit{i.e.} in the half-plane $x^1<0$). The boundary extends to the entire $x^2$ space.
This derivation is based on the assumption that the metric depends only on $x^2$ near the boundary. Then $x^1$ and $x^2$ are completely decoupled in the boundary Hamiltonian. 
The wave function of the boundary mode of the Hamiltonian $\mathcal{H}_0$ is the product of a plane wave of the $x^2$ coordinate and a two-component spinor wave function of the $x^1$ coordinate.
\be
\psi(x^1,x^2)=\psi_2(x^2)\psi_1(x^1).
\ee
Based on the discussion in Sect.~\ref{sec:edge_modes}, the $k=0$ approximation~\cite{qi2011topological} for the wave function is sufficiently accurate for this system. Therefore, the boundary modes satisfy the following equation:
\be \label{eq:d_boun_equ}
\Big(-i\gamma^{\hat{0}}\gamma^{\hat{1}}\partial_{1}+m\gamma^{\hat{0}}-\gamma^{\hat{0}}\partial_1^2
\Big)\psi_1(x^1)&=&0
\ee
which, combined with the open boundary condition $\psi_1(0)=0$, gives the solution
\be \label{eq:bound_wave_function}
\psi_1(x^1)&=&\frac{1}{\sqrt{N}}(e^{-\lambda_1x}-e^{-\lambda_2x})\vert s\rangle\equiv \psi_0(x^1)\vert s\rangle
\qquad \text{with}\qquad \lambda_{1,2}=\frac{1}{2}\Big(1\pm \sqrt{1+4m}
\Big).
\ee
The two-component spinor $\vert s\rangle$ corresponding to the edge bound states satisfies $\sigma_y\vert s\rangle= -\vert s\rangle$ or $i\gamma^{\hat{1}}\vert s\rangle =\vert s\rangle$. The boundary Hamiltonian for the bulk Hamiltonian $H$ is
\be
I&=&\langle \psi_1\vert\sqrt{g}\gamma^{\hat{0}}(-\frac{i}{2}e^{j}_{*\alpha}\gamma^{\alpha}\rpartial_j+\frac{i}{2}\lpartial_j\gamma^{\alpha}e^j_{*\alpha}+m-\eta^{ab}e^{i}_{*a}e^{j}_{*b}\lpartial_i\rpartial_j)\vert\psi_1\rangle
\nn
&=&\langle \psi_1\vert\sqrt{g}( -\frac{i}{2}e^{j}_{*\hat{2}}\gamma^{\hat{0}}\gamma^{\hat{2}}\rpartial_j+\frac{i}{2}\lpartial_j\gamma^{\hat{0}}\gamma^{\hat{2}}e^{j}_{*\hat{2}}-\frac{i}{2}e^{j}_{*\hat{0}}\gamma^{\hat{0}}\gamma^{\hat{0}}\rpartial_j+\frac{i}{2}\lpartial_j\gamma^{\hat{0}}\gamma^{\hat{0}}e^{j}_{*\hat{0}})
\vert \psi_1\rangle
\nn
&=&\sqrt{g}(e^{2}_{*\hat{2}}-e^{2}_{*\hat{0}})\frac{i}{2}\Big(\rpartial_2-\lpartial_2
\Big)
\ee
Here we used the following identities (remember that the metric depends only on $x^2$ near the boundary.)
\be \label{eq:d_x1_equ}
\int dx^1 \psi_0^{*}(x^1)\partial_1\psi_0(x^1)=0
\ee
Since the wave function $\psi_1(x^1)$ in Eq.~(\ref{eq:bound_wave_function}) is either a real function or a purely imaginary function.
\be
II&=&\langle\psi_1 \vert \sqrt{g}\gamma^{\hat{0}}\frac{i}{2}e^{j}_{*\alpha}(\gamma^{\alpha}\omega_j+\omega_j\gamma^{\alpha})\vert \psi_1\rangle
=-\frac{i}{8}\sqrt{g}e^{j}_{*\alpha}\langle s\vert \gamma^{\hat{0}}\gamma^{\alpha}\gamma^{x}\gamma^{y}+\gamma^{\hat{0}}\gamma^{x}\gamma^{y}\gamma^{\alpha}\vert s\rangle \omega_{j xy}=0
\ee
The reasoning here is as follows: for any $x\ne y$, if $\alpha$ is not equal to either $x$ or $y$, then the structure inside the matrix is $\langle s\vert \gamma^{0}\gamma^{0}\gamma^{1}\gamma^{2}\vert s\rangle$, which is clearly zero. If $\alpha$ equals either $x$ or $y$, then by the properties of the Clifford algebra, it is evident that the two terms inside the matrix will cancel each other out.
\be \label{eq:d_bound_3}
III&=&-\eta^{ab}e^{i}_{*a}e^{j}_{*b}\langle \psi_1\vert \sqrt{g}\gamma^{\hat{0}}\Big(-\lpartial_i\omega_j+\omega_i\rpartial_j-\omega_i\omega_j
\Big)\vert \psi_1\rangle=\eta^{ab}e^{i}_{*a}e^{j}_{*b}\langle \psi_1\vert \sqrt{g}\gamma^{\hat{0}}\omega_i\omega_j\vert \psi_1\rangle
\nn
&=&\frac{1}{2}\eta^{ab}\sqrt{g}e^{i}_{*a}e^{j}_{*b}\Big(\langle s\vert\gamma^{\hat{0}}\omega_i\omega_j\vert s\rangle +\langle s\vert \gamma^{\hat{0}}\omega_j\omega_i\vert s\rangle 
\Big) =0
\ee
The argument for Eq.~(\ref{eq:d_bound_3}) is as follows: In the first equality of the above equation, we recognize that its first two terms can only contribute terms such as $\omega_{i\hat{0}\hat{1}}$ or $\omega_{i\hat{1}\hat{2}}$. However, these terms do not contribute to the energy current $j^E_2$~(\ref{eq:d_enegy_curr_deriv}). Moreover, the vielbein field part also does not contribute to the energy current~(Eq. (\ref{eq:d_enegy_curr_deriv}) and Eq.~(\ref{eq:d_deriv})). Therefore, we can omit this term.
In the second line of Eq.~(\ref{eq:d_bound_3}), we utilize the symmetry of the metric tensor $\eta^{ab}=\eta^{ba}$, which allows us to express the third term in Eq.~(\ref{eq:d_bound_3}) in a symmetric form. The key point here is that the components of the spin connection can always be represented by a single gamma matrix, for example $\gamma^{\hat{0}}\gamma^{\hat{1}}\omega_{i\hat{0}\hat{1}} \sim \gamma^{\hat{2}}\omega_{i\hat{0}\hat{1}}$. When the gamma matrices corresponding to the two spin connections are the same, the third term in Eq.~(\ref{eq:d_bound_3}) is proportional to $\langle s\vert \gamma^{\hat{0}}\vert s\rangle$, which is clearly zero. When the gamma matrices of the two spin connections are different, the symmetric form ensures that the two terms cancel each other out due to the properties of the Clifford algebra.
\be
\mathcal{H}=\langle\psi_1\vert H\vert \psi_1\rangle=I+I+III=\sqrt{g}(e^{2}_{*\hat{2}}-e^{2}_{*\hat{0}})\frac{i}{2}\Big(\rpartial_2-\lpartial_2\Big)
\ee

Now we consider the source term $H_{\rm src}$,
\be
I_{\rm src}&=&i\langle \psi_1\vert e^{i}_{*\hat{0}}\lpartial_i\phi_g(-i\gamma^j\rpartial_j+m-\rpartial^2)\vert \psi_1\rangle-i\langle \psi_1\vert e^{i}_{*\hat{0}}\phi_g(-i\lpartial_j\gamma^j+m-\lpartial^2)\rpartial_i\vert \psi_1\rangle
\nn
&\overset{(\ref{eq:d_boun_equ})}{=}&i\langle \psi_1\vert e^{i}_{*\hat{0}}\lpartial_i\phi_g(-i\gamma^{\hat{2}}\rpartial_2-\rpartial_2^2)\vert \psi_1\rangle-i\langle \psi_1\vert e^{i}_{*\hat{0}}\phi_g(-i\lpartial_2\gamma^{\hat{2}}-\lpartial^{2}_2)\rpartial_i\vert \psi_1\rangle
\nn
&\overset{(\ref{eq:d_x1_equ})}{=}&ie^{2}_{*\hat{0}}\phi_g\langle s\vert \lpartial_2(-i\gamma^{\hat{2}}\rpartial_2-\rpartial_2^2)-(-i\lpartial_2\gamma^{\hat{2}}-\lpartial^{2}_2)\rpartial_2\vert s\rangle=ie^{2}_{*\hat{0}}\phi_g\Big(-\lpartial_2\rpartial_2^2+\lpartial_2^2\rpartial_2
\Big)
\nn
II_{\rm src}&=&i\langle \psi_1\vert e^{i}_{*\hat{0}}\Big(\lpartial_i\frac{-i}{2}(\partial_j\phi_g)\gamma^j+\frac{i}{2}(\partial_j\phi_g)\gamma^j\rpartial_i
\Big)\vert \psi_1\rangle
\nn
&=&i\langle \psi_1\vert e^{i}_{*\hat{0}}\Big(\lpartial_i\frac{-i}{2}(\partial_2\phi_g)\gamma^{\hat{2}}+\frac{i}{2}(\partial_2\phi_g)\gamma^{\hat{2}}\rpartial_i
\Big)\vert \psi_1\rangle=0
\nn
III_{\rm src}&=&i\langle \psi_1\vert e^{i}_{*\hat{0}}\Big(-\lpartial_i(\partial_j\phi_g)\rpartial_j+\lpartial_j(\partial_j\phi_g)\rpartial_i
\Big)\vert \psi_1\rangle
=i(\partial_2\phi_g)e^{i}_{*\hat{0}}\langle \psi_1\vert -\lpartial_i\rpartial_2+\lpartial_2\rpartial_i\vert \psi_1\rangle
\nn
&=&i(\partial_2\phi_g)e^{1}_{*\hat{0}}\langle \psi_1\vert -\lpartial_1\rpartial_2+\lpartial_2\rpartial_1\vert \psi_1\rangle\overset{(\ref{eq:d_x1_equ})}{=}0
\ee
In the derivation of the term $II_{\rm src}$, we relied on the assumption that the metric depends only on $x_2$. In summary,
\be
\mathcal{H}_{\rm src}&=&\langle \psi_1\vert H_{\rm src}\vert \psi_1\rangle=I_{\rm src}+II_{\rm src}+III_{\rm src}=ie^{2}_{*\hat{0}}\phi_g\Big(-\lpartial_2\rpartial_2^2+\lpartial_2^2\rpartial_2
\Big)
\nn
\mathcal{H}_{\rm ex}&=&\sqrt{g}(e^{2}_{*\hat{2}}-e^{2}_{*\hat{0}})\frac{i}{2}\Big(\rpartial_2-\lpartial_2\Big)+ie^{2}_{*\hat{0}}\phi_g\Big(-\lpartial_2\rpartial_2^2+\lpartial_2^2\rpartial_2
\Big)
\ee
and we can express the boundary Hamiltonian as
\be
\mathcal{H}_{\rm ex}&=&\int dx^2\psi^{\dagger}_2(x^2)\Big(\frac{i}{2}(\rpartial_2-\lpartial_2)
+\frac{i}{2}\zeta(x^2)(\rpartial_2-\lpartial_2)+i\chi(x^2)(-\lpartial_2\rpartial_2^2+\lpartial_2^2\rpartial_2)\Big)\psi_2(x^2)
\nn
\zeta(x^2)&=&\sqrt{g}(e^{2}_{*\hat{2}}-e^{2}_{*\hat{0}})-1=(1+\frac{h}{2})(1-\frac{1}{2}h^{2}_{*\hat{2}}+\frac{1}{2}h^{2}_{*\hat{0}})-1=\frac{h}{2}+(1+\frac{h}{2})(\frac{1}{2}h^{2}_{*\hat{0}}-\frac{1}{2}h^{2}_{*\hat{2}})
\nn
\chi(x^2)&=&\phi_g e^{2}_{*\hat{0}}=\frac{1}{2}(1+\phi)(-h^{2}_{*\hat{0}})
\ee
with 
\be
e^{*\alpha}_{\mu}=\delta_{\mu}^{*\alpha}+h^{*\alpha}_{\mu}/2,\quad e^{\mu}_{*\alpha}=\delta^{\mu}_{*\alpha}-h^{\mu}_{*\alpha}/2,\quad \sqrt{g}=1+h/2
\ee
And we have the following identity:
\be \label{eq:d_deriv_2}
\frac{\delta \zeta (x^2)}{\delta e^{*\hat{2}}_0(x^2)}=1,\qquad \frac{\delta \chi(x^2)}{\delta e^{*\hat{2}}_0(x^2)}=-1
\ee
Observing the functions $\zeta(x^2)$ and $\chi(x^2)$, we find that when the metric tensor is sufficiently close to flat spacetime, $\zeta$ and $\chi$ both become sufficiently small. Therefore, these terms can be treated as perturbative corrections. In the next section, we will take this as the starting point for performing a step-by-step calculation using perturbation theory at finite temperature $1/\beta$.

\section{Boundary free energy and Boundary current} \label{d_Boundary_free_energy_and_Boundary_current}
In this section, we calculate the effective free energy of the vielbein field for the boundary modes. We begin by presenting the expression for the partition function in this context.
\be
Z=\int \mathcal{D}[\psi^{*},\psi]\exp\Big(-\mathcal{S}[\psi^{*},\psi,\zeta,\chi]
\Big)=\int \mathcal{D}[\psi^*,\psi] \exp\Big(-\mathcal{S}_0[\psi^{*},\psi]-\mathcal{S}_1[\psi^{*},\psi,\zeta,\chi]\Big)
\ee
with
\be \label{eq:d_bound_action}
\mathcal{S}_0[\psi^*,\psi]&=&\int^{\beta}_{0}d\tau \int dx\psi^*(x,\tau)\Big(\partial_{\tau}+\frac{i}{2}\rpartial_x-\frac{i}{2}\lpartial_x
\Big)\psi(x,\tau)=\sum_{\omega_n,k}\psi^*_{n,k}(-i\omega_n-k)\psi_{n,k}
\nn
\mathcal{S}_1[\psi^{*},\psi,\zeta,\chi]&=&\int_0^{\beta}d\tau \int dx\psi^*(x,\tau)\Big(\frac{i}{2}\zeta(x)\rpartial_x-\frac{i}{2}\lpartial_x\zeta(x)+i\lpartial_x^2\chi(x)\rpartial_x-i\lpartial_x\chi(x)\rpartial_x^2
\Big)\psi(x,\tau)
\nn
&=&\sum_{n,k,k'}\psi^*_{n,k}\Big(-\frac{k'}{2}\zeta(k-k')-\frac{k}{2}\zeta(k-k')+(-k^2)\chi(k-k')(-k')-(k)\chi(k-k')(-k'^2)
\Big)\psi_{n,k'}
\nn
&=&\sum_{n,k,k'}\psi^*_{n,k}\Big(-\frac{k+k'}{2}\zeta(k-k')+(k^2k'+kk'^2)\chi(k-k')\Big)\psi_{n,k'}
\ee
and the expansion of the boundary free energy is given by~\cite{stoof2009ultracold,nakai2016finite}
\be \label{eq:d_bound_free_energy_expan}
\mathcal{F}[\zeta,\chi]=\frac{1}{\beta}\sum_{l=1}^{\infty}\frac{1}{l}\text{Tr}[(G_0\Sigma)^l]
\ee
According to Eq.~(\ref{eq:d_bound_action}), the inverse Green function and the self-energy of the system can be expressed in the following form:
\be \label{eq:d_bound_green}
G^{-1}_0(k,\omega_n)&=&i\omega_n+k,\quad 
\Sigma(k,k')=-\frac{k+k'}{2}\zeta(k-k')+(k^2k'+kk'^2)\chi(k-k')
\ee
Combining Eq.~(\ref{eq:d_bound_green}), we now proceed to calculate each term in the expansion of Eq.~(\ref{eq:d_bound_free_energy_expan}). The first-order perturbation term is given by:
\be
\frac{1}{\beta}\text{Tr}[G_0\Sigma]&=&\frac{1}{\beta}\sum_{\omega_n}\int \frac{dp}{2\pi}\Sigma(p,p)G_0(p,\omega_n)=\frac{1}{\beta}\sum_{\omega_n}\int \frac{d p}{2\pi}(-p\zeta(0)+2p^3\chi(0))G_0(p,\omega_n)
\nn
&=&\frac{1}{\beta}\int dx\zeta(x)\sum_{\omega_n}\int \frac{dp}{2\pi}\frac{-p}{i\omega_n+p}+\frac{1}{\beta}\int dx \chi(x)\sum_{\omega_n}\int \frac{dp}{2\pi} \frac{2p^3}{i\omega_n+p}
\ee
here,
\be
\sum_{\omega_n}\int \frac{dp}{2\pi}\frac{-p}{i\omega_n+p}&=&\sum_{\omega_n}\int \frac{dp}{2\pi}\frac{p}{i\omega_n-p}=\beta\int \frac{dp}{2\pi} p f(p)
=\lim_{\Lambda\rightarrow \infty}\beta \int^{\Lambda}_{-\Lambda}\frac{dp}{2\pi}pf(p)=\lim_{\Lambda\rightarrow \infty}\frac{1}{2\pi \beta}\int^{\beta \Lambda}_{-\beta\Lambda}dx\frac{x}{e^x+1}
\nn
&=&\frac{1}{2\pi\beta}\frac{-\beta^2\Lambda^2}{2}-\frac{1}{2\pi\beta}\int dx \frac{x^2}{2}\frac{-e^x}{(1+e^x)^2}=-\frac{\beta}{4\pi}\Lambda^2+\beta \frac{\pi T^2}{12}
\nn
\sum_{\omega_n}\int \frac{dp}{2\pi}\frac{2p^3}{i\omega_n+p}&=&-2\sum_{\omega_n}\int \frac{dp}{2\pi}\frac{p^3}{i\omega_n-p}=-\frac{2\beta}{2\pi}\lim_{\Lambda\rightarrow \infty} \int_{-\Lambda}^{\Lambda} dp p^3 f(p)=-\frac{1}{\pi\beta^3}\lim_{\Lambda\rightarrow \infty}\int^{\beta \Lambda}_{-\beta \Lambda}dx \frac{x^3}{e^x+1}
\nn
&=&-\frac{1}{\pi\beta^3}\Big(-\frac{\beta^4\Lambda^4}{4}-\int dx\frac{x^4}{4}\frac{-e^x}{(1+e^x)^2} 
\Big)=\beta \frac{\Lambda^4}{4\pi}-\beta \frac{7\pi^3 T^4}{60}
\ee
In the derivation of this expression, we utilize the identity:
\be
\int dx \frac{x}{e^x+1}=\int dx \Big(\frac{x^2}{2}\frac{1}{e^x+1}
\Big)'-\int dx \frac{x^2}{2}\Big(\frac{1}{e^x+1}\Big)'
\ee
and the first-order term is 
\be
\frac{1}{\beta}\text{Tr}[G_0\Sigma]&=&\Big(-\frac{1}{4\pi}\Lambda^2 + \frac{\pi T^2}{12}\Big)\int dx \zeta(x)+\Big( \frac{\Lambda^4}{4\pi}- \frac{7\pi^3 T^4}{60}\Big)\int dx \chi(x)
\nn
&=&\frac{\pi T^2}{12}\int dx \zeta(x)-\frac{7\pi^3 T^4}{60}\int dx \chi(x)
\ee
In this expression, we have omitted nonphysical terms that are independent of temperature. Next, using a similar method, we calculate the second-order perturbation term: 
\be \label{eq:d_free_second}
\frac{1}{2\beta}\text{Tr}[(G_0\Sigma)^2]&=&\frac{1}{2\beta}\sum_{\omega_n}\int \frac{dp_1}{2\pi}\frac{dp_2}{2\pi}G_0(p_1,\omega_n)\Sigma(p_1,p_2)G_0(p_2,\omega_n)\Sigma(p_2,p_1)
\nn
&=&\frac{1}{2\beta}\sum_{\omega_n}\int \frac{dp_1dp_2}{4\pi^2}\Big(\frac{(p_1+p_2)^2}{4}\zeta(p_1-p_2)\zeta(p_2-p_1)+\frac{-(p_1+p_2)}{2}(p_1^2p_2+p_1p_2^2)\zeta(p_1-p_2)\chi(p_2-p_1)
\nn
&&+\frac{-(p_1+p_2)}{2}(p_1^2p_2+p_1p_2^2)\chi(p_1-p_2)\zeta(p_2-p_1)\Big)G_0(p_1,\omega_n)G_0(p_2,\omega_n)
\nn
&=&\frac{1}{2\beta}\sum_{\omega_n}\int \frac{dpdq}{4\pi^2}\Big(\frac{(2p+q)^2}{4}\zeta(q)\zeta(-q)+\frac{-(2p+q)}{2}((p+q)^2p+(p+q)p^2)\zeta(q)\chi(-q)
\nn
&&+\frac{-(2p+q)}{2}((p+q)^2p+(p+q)p^2)\chi(q)\zeta(-q)\Big)G_0(p+q,\omega_n)G_0(p,\omega_n)
\nn
&=&\frac{1}{2\beta}\int \frac{dq}{2\pi}\int dxdye^{iq(x-y)}\Big(\Pi^{(1)}(q)\zeta(x)\zeta(y)+\Pi^{(2)}(q)\zeta(x)\chi(y)+\Pi^{(2)}(q)\chi(x)\zeta(y)\Big)
\ee
with 
\be
\Pi^{(1)}(q)&=&\sum_{\omega_n}\int \frac{dp}{2\pi}\frac{(2p+q)^2}{4}G_0(p+q,\omega_n)G_0(p,\omega_n)
\nn
\Pi^{(2)}(q)&=&\sum_{\omega_n}\int \frac{dp}{2\pi}\frac{-(2p+q)}{2}((p+q)^2p+(p+q)p^2)G_0(p+q,\omega_n)G_0(p,\omega_n)
\ee
In the derivation of Eq.~(\ref{eq:d_free_second}), we have neglected the quadratic terms in
$\chi(x)$. This is because, in the Luttinger case, it is strictly zero. As a result, if a term in the expression of free energy contains more than one $\chi(x)$ function, its contribution to the energy current will necessarily be zero.
Besides, here we only need to consider $\Pi^{(1,2)}(q)$ at $q=0$, as it is clear that expanding $\Pi^{(1,2)}(q)$ in powers of $q$ results in higher-order terms that are proportional to higher-order derivatives of $\zeta(x)$ or $\chi(x)$. Since the primary focus is on contributions to the free energy that do not involve such derivatives, we can neglect these terms. Consequently, we will set $q=0$ in $\Pi^{(1,2)}(q)$ for the subsequent analysis.
\be
\Pi^{(1)}(0)&=&\sum_{\omega_n}\int \frac{dp}{2\pi}\frac{p^2}{(i\omega_n+p)^2}=\sum_{\omega_n}\int \frac{dp}{2\pi}\frac{p^2}{(i\omega_n-p)^2}=\frac{\beta}{2\pi} \int  dp \frac{df(p)}{dp}p^2
\nn
&=&\frac{1}{2\pi\beta }\int dx x^2 \frac{-e^x}{(1+e^x)^2}=-\frac{\pi}{6\beta}=-\beta\frac{\pi T^2}{6}
\nn
\Pi^{(2)}(0)&=&\sum_{\omega_n}\int \frac{dp}{2\pi}\frac{-p(2p^3)}{(i\omega_n+p)^2}=-\sum_{\omega_n}\int \frac{dp}{2\pi}\frac{2p^4}{(i\omega_n-p)^2}=-\frac{2\beta}{2\pi}\int dp \frac{df(p)}{dp}p^4
\nn
&=&-\frac{1}{\pi\beta^3}\int dx x^4\frac{-e^x}{(1+e^x)^2}=\frac{1}{\pi\beta^3}\frac{7\pi^4}{15}=\beta \frac{7\pi^3 T^4}{15}
\ee
and
\be
\frac{1}{2\beta}\text{Tr}[(G_0\Sigma)^2]&=&\frac{1}{2\beta}\int \frac{dq}{2\pi}\int dxdye^{iq(x-y)}\Big(\Pi^{(1)}(0)\zeta(x)\zeta(y)+\Pi^{(2)}(0)\zeta(x)\chi(y)+\Pi^{(2)}(0)\chi(x)\zeta(y)\Big)
\nn
&=&-\frac{\pi T^2}{12}\int dx (\zeta(x))^2+\frac{7\pi^3T^4}{15}\int dx \zeta(x)\chi(x)
\ee
In general, the $l$-th order term of the expansion can be represented as
\be
\frac{1}{l\beta}\text{Tr}[(G_0\Sigma)^l]&=&\frac{1}{l\beta}U_l^{(1)}\int dx (\zeta(x))^l+\frac{1}{\beta}U_l^{(2)}\int dx (\zeta(x))^{l-1}\chi(x)
\nn
&=&(-1)^{l+1}\frac{\pi T^2}{12}\int dx (\zeta(x))^l+(-1)^l\frac{(l+2)(l+1)l\pi^3T^4}{360}\int dx (\zeta(x))^{l-1}\chi(x)
\ee
with Hertz’s approach, we convert the above equation into a contour integral and perform the Matsubara sum, which gives the following equation:
\be
U^{(1)}_{l}&=&\sum_{\omega_n}\int \frac{dp}{2\pi}\frac{(-p)^l}{(i\omega_n+p)^l}=\sum_{\omega_n}\int \frac{dp}{2\pi}\frac{p^l}{(i\omega_n-p)^l}=\frac{\beta}{(l-1)!}\int \frac{dp}{2\pi}p^l[f(p)]^{(l-1)}
\nn
&=&(-1)^{l-2}\frac{l!\beta}{2(l-1)!}\int \frac{dp}{2\pi}p^2\frac{df(p)}{dp}=(-1)^l\frac{l}{4\pi\beta}\int dx x^2\frac{-e^x}{(1+e^x)^2}=(-1)^{l+1}l\beta \frac{\pi T^2}{12}
\nn
U^{(2)}_l&=&\sum_{\omega_n}\int \frac{dp}{2\pi}\frac{(-p)^{l-1}2p^3}{(i\omega_n+p)^l}=-2\sum_{\omega_n}\int \frac{dp}{2\pi}\frac{p^{l+2}}{(i\omega_n-p)^l}=
-2\frac{\beta}{(l-1)!}\int \frac{dp}{2\pi}p^{l+2}[f(p)]^{l-1}
\nn
&=&-2(-1)^{l-2}\frac{\beta(l+2)!}{24(l-1)!}\int \frac{dp}{2\pi}p^4\frac{df(p)}{dp}=(-1)^{l+1}\frac{7(l+2)(l+1)l}{24\pi\beta^3}\int dx x^4\frac{-e^x}{(1+e^x)^2}
\nn
&=&\beta(-1)^l\frac{(l+2)(l+1)l\pi^3T^4}{360}
\ee
By summing these contributions together, we obtain the following complete expression for the boundary effective free energy:
\be \label{eq:d_free_energy_final}
\mathcal{F}[\zeta,\chi]&=&\frac{\pi T^2}{12}\int dx \Big(\sum_{l=1}(-1)^{l+1}(\zeta(x))^l
\Big)+\frac{7\pi^3 T^4}{360}\int dx \chi(x)\Big(\sum_{l=1}(-1)^l(l+2)(l+1)l (\zeta(x))^{l-1}
\Big)
\nn
&=&\frac{\pi T^2}{12}\int dx \frac{\zeta(x)}{1+\zeta(x)}-\frac{7\pi^3 T^4}{60}\int dx \frac{\chi(x)}{(1+\zeta(x))^4}
\ee
By combining Eqs.~(\ref{eq:d_enegy_curr_free_deriv}), (\ref{eq:d_deriv}), (\ref{eq:d_deriv_2}), and (\ref{eq:d_free_energy_final}), we obtain the expression for the boundary energy current as follows:
\be \label{eq:d_bound_current_result}
j^{2,\text{bdry}}_E(x^2)&=&-(1+\phi)^2\frac{\delta \mathcal{F}}{\delta e^{*\hat{2}}_{0}(x^2)}-\sum_{i=1,2}(1+\phi(x^2))\partial_i\Big(\frac{\delta \mathcal{F}}{\delta \omega_{i\hat{0}\hat{2}}(x^2)}
\Big)=-(1+\phi(x^2))^2\frac{\delta \mathcal{F}}{\delta e^{*\hat{2}}_{0}(x^2)}
\nn
&=&-(1+\phi(x^2))^2\frac{\pi T^2}{12}\frac{1}{(1+\zeta(x))^2}\frac{\delta \zeta(x^2)}{\delta e^{*\hat{2}}_0(x^2)}+(1+\phi(x^2))^2\frac{7\pi^3 T^4}{60}\frac{1}{(1+\zeta(x^2))^4}\frac{\delta \chi(x^2)}{\delta e^{*\hat{2}}_0(x^2)}
\nn
&=&-\frac{\pi T^2}{12}-\frac{7\pi^3 T^4}{60}\frac{1}{(1+\phi(x^2))^2}
\ee

\section{Energy magnetization and thermal Hall conductivity}
In this section, based on the results derived from Eq.~(\ref{eq:d_bound_current_result}), we calculate the system's energy magnetization and thermal Hall conductivity.
The energy magnetization, denoted as $\mathbf{M}_E$, is defined as 
\be
\mathbf{j}_E=\nabla\times \mathbf{M}_E.
\ee
Under these conditions, the bulk energy magnetization $M_E^z$ can be expressed in terms of the energy current at the boundary as follows:
\be
j^{2,\text{bdry}}_E&\equiv& \int_0^{+\infty} dx^1 j^2_E(x^1)=\int_0^{+\infty}dx^1(-\partial_1 M_E^z(x^1))=-M^z_E(+\infty)+M_E^{z}(0)=-M_E^z(+\infty)\equiv -M_E^z
\ee
and
\be
M_E^z&\equiv& M^E_z(+\infty)=\frac{\pi T^2}{12}+\frac{7\pi^3 T^4}{60}
\ee
In the above derivation, we used the condition that $M_E^z$ equals zero at the boundary~\cite{zhang2020thermodynamics}. Furthermore, since $M_E^z$  is a quantity defined in equilibrium, we only need to consider a uniform gravitational potential $\phi$ that is identically. 
Note that, under these conditions, the approximations made in Section \ref{d_Boundary_free_energy_and_Boundary_current} hold rigorously.

In addition, since the bulk energy current $j_E$ near the boundary satisfies the following equation:
\be
j_E^1(x^1=+0)=-\partial_2j_E^{\rm bdry}(x^2).
\ee
The bulk energy current associated with the boundary energy current Eq.~(\ref{eq:d_bound_current_result}) is expressed as follows:
\be \label{qe:d_bulk_current}
j^1_{E}(x^1=+0)=-\frac{7\pi^3T^4}{30}\frac{1}{(1+\phi(x^2))^3}\partial_2\phi(x^2)=-\frac{7\pi^3T^4}{30}\partial_2\phi(x^2)
\ee
In the position of the last equality in Eq.~(\ref{qe:d_bulk_current}), we have applied a linear approximation to $\phi$. Furthermore, in Section \ref{d_Boundary_free_energy_and_Boundary_current}, we also made a linear approximation for $\partial \phi$. This implies that the general expression for $j_E^1$ takes the form 
\be
j^1_{E}(x^1=+0)=-\frac{7\pi^3T^4}{30}\partial_2\phi(x^2)+f_1(\phi^2,\phi^3,\cdots)\partial_2\phi(x^2)+f_2(\partial^2\phi,\partial^3\phi,\cdots;(\partial\phi)^2,(\partial\phi)^2,\cdots)
\ee
By comparing with equation~(\ref{qe:d_bulk_current}), it can be observed that the additional two terms do not contribute to the calculation of linear response coefficients such as the thermal Hall conductivity.
According to linear response theory and the Tolman-Ehrenfest relation~\cite{luttinger1964theory,qin2011energy}, the thermal conductivity contributed by the quantum anomaly is
\be
\kappa^{\text{Bulk}}_{xy}=\frac{7\pi^3 T^3}{30}
\ee
Combining with the contribution from the energy magnetization~\cite{guo2020gauge}, the thermal Hall conductivity is then given by
\be
\kappa_{xy}=\kappa^{\text{Bulk}}_{xy}+\frac{2M^z_E}{T}=\frac{\pi T}{6}+\frac{7\pi^3T^3}{15}
\ee
It is worth noting that this result is identical to the one obtained directly using the generalized Streda formula:
\be
\kappa_{xy}=\frac{\partial M_E^z}{\partial T}=\frac{\pi T}{6}+\frac{7\pi^3T^3}{15}
\ee

\section{Edge modes of Topological Insulator} \label{sec:edge_modes}
We start with a standard continuum model describing a topological insulator,
\be
\hat{h}&=&\hat{k}_x\sigma_x+\hat{k}_y\sigma_y+(m+\hat{k}_x^2+\hat{k}_y^2)\sigma_z.
\ee
Consider the model Hamiltonian defined on the half-space $x>0$ in the $x-y$ plane. We can divide the one-body Hamiltonian $\hat{h}$ into two parts, $\hat{h}=\hat{h}_0+\hat{h}_1$,
\be
\hat{h}_0&=&\hat{k}_x\sigma_x+(m+\hat{k}_x^2)\sigma_z,
\nn
\hat{h}_1&=&\hat{k}_y\sigma_y+\hat{k}_y^2\sigma_z
\ee
All $k_x$-dependent terms are include in $\hat{h}_0$. For such a semi-infinite system, $k_y$ is a good quantum number, but $k_x$ must be replaced by the operator $-i\partial_x$.
\be
\left(\begin{matrix}
m-\partial_x^2+k_y^2&& -i\partial_x-ik_y
\\
-i\partial_x+ik_y&& -m+\partial_x^2-k_y^2
\end{matrix}
\right)\psi_{k_y}(x)=E\psi_{k_y}(x)
\ee
With the wave function ansatz $\psi_{k_y}=e^{\lambda x}\vert s\rangle=e^{\lambda x}(\psi_u,\psi_d)^T$, this Schrodinger equation can be simplified to
\be
\left(\begin{matrix}
m-\lambda^2+k_y^2&& -i\lambda-ik_y
\\
-i\lambda+ik_y&& -m+\lambda^2-k_y^2
\end{matrix}
\right)\vert s\rangle &=&E\vert s\rangle
\nn
(m-\lambda^2+k_y^2)\psi_u-i(\lambda+k_y)\psi_d&=&E\psi_u
\nn
-i(\lambda-k_y)\psi_u-(m-\lambda^2+k_y^2)\psi_d&=&E\psi_d
\ee
This equation gives four roots $\pm \lambda_1$, $\pm\lambda_2$
\be
\lambda_{1}^2&=&k_y^{2}+m+\frac{1}{2}+ \frac{1}{2}\sqrt{1+4E^2+4m},\quad
\psi_{d,1\pm}=-i\frac{\pm\lambda_{1}-k_y}{E+(m-\lambda_{1}^2+k_y^2)}\psi_u
\nn
\lambda_{2}^2&=&k_y^{2}+m+\frac{1}{2}- \frac{1}{2}\sqrt{1+4E^2+4m},
\quad
\psi_{d,2\pm}=-i\frac{\pm\lambda_2-k_y}{E+(m-\lambda_2^2+k_y^2)}\psi_u
\ee
Therefore, the general solution is given by
\be
\psi_{ky}=ae^{\lambda_1 x}\psi_{1+}+be^{-\lambda_1x}\psi_{1,-}+ce^{\lambda_2 x}\psi_{2,+}+de^{-\lambda_2 x}\psi_{2,-}
\ee
To make this expression clearer, we restrict our consideration to the special case of $k_y=0$. Due to the one-body Hamiltonian, the wave function $\psi_{k_y=0}$ has particle-hole symmetry. Therefore, we expect that a special edge state $\psi_{k_y}=0$ with $E=0$ can exist.
\be
\lambda_1^2&=&m+\frac{1}{2}+\frac{1}{2}\sqrt{1+4m}=\Big(\frac{1}{2}(1+\sqrt{1+4m})\Big)^2
\nn
\lambda_2^2&=&m+\frac{1}{2}-\frac{1}{2}\sqrt{1+4m}=\Big(\frac{1}{2}(1-\sqrt{1+4m})\Big)^2
\ee
and the two-component spinor becomes 
\be
\psi_{d,1\pm}&=&\pm i\psi_u, \quad \psi_{d,2\pm}=\pm i\psi_u
\ee
Now we return to the general case. The coefficients $a,b,c,d$ can be determined by imposing the open boundary condition $\psi_{k_y}(0)=0$ and the normalizability of the wave function in the region $x>0$.

Let us first consider the constraint on the wave function imposed by normalizability: $\text{Re}\lambda_{1,2}<0$ ($b=d=0$) or $\text{Re}\lambda_{1,2}>0$ ($a=c=0$). As seen from the formalism of $\lambda_{1,2}$ for zero modes, these conditions can be satisfied only in the regime when $m<0$. The wave function becomes 
\be
\psi_{k_y}=be^{-\lambda_1x}\psi_{1,-}+de^{-\lambda_2 x}\psi_{2,-}
\ee
Now, we consider the impact of the boundary conditions, which essentially impose the following two constraints on the wave function: $d=-b$ and $\psi_{d,1-}=\psi_{d,2-}$. Using the second condition, we can derive the energy dispersions for electrons.
\be
&&\frac{-\lambda_{1}-k_y}{E+(m-\lambda_{1}^2+k_y^2)}=\frac{-\lambda_2-k_y}{E+(m-\lambda_2^2+k_y^2)}
\rightarrow E=-k_y(\lambda_1+\lambda_2)-k_y^2-m-\lambda_1\lambda_2
\ee
This equation is difficult to solve exactly; however, when we restrict our consideration to small $k_y$, we can compute its derivatives step by step. First, based on the particle hole symmetry, we know that $E(k_y=0)=0$. Using this as a starting point, we proceed to compute its first derivative.
\be
\frac{dE}{dk_y}\Big\vert_{k_y=0}&=&\Big[-(\lambda_1(0)+\lambda_2(0))-k_y(\frac{d\lambda_1}{dk_y}+\frac{d\lambda_2}{d k_y})-2k_y-\frac{d\lambda_1}{d k_y}\lambda_2-\lambda_1\frac{d\lambda_2}{d k_y}\Big]_{k_y=0}=-1
\nn
\frac{d\lambda_1}{dk_y}\Big\vert_{k_y=0}&=&\frac{E(0)}{\sqrt{1+4m+4E(0)^2}\sqrt{\frac{1}{2}+m+\frac{1}{2}\sqrt{1+4m+4E(0)^2}}}\frac{dE}{dk_y}\Big\vert_{k_y=0}=0
\ee
Now that we have obtained its first derivative, we can use it to compute the second derivative. Continuing this process iteratively, we arrive at the following series of results.
\be
\frac{d^2E}{dk_y^2}\Big\vert_{k_y=0}&=&-(\frac{d\lambda_1}{dk_y}+\frac{d\lambda_2}{d k_y})-2-2\frac{d\lambda_1}{dk_y}\frac{d\lambda_2}{dk_y}-\lambda_1\frac{d^2\lambda_2}{dk_y^2}-\frac{d^2\lambda_1}{dk_y^2}\lambda_2=-2-\lambda_1\frac{d^2\lambda_2}{dk_y^2}-\frac{d^2\lambda_1}{dk_y^2}\lambda_2
\nn
&=&-2-\frac{2\sqrt{2}m}{\sqrt{1+4m}}\Big(\frac{1}{\sqrt{1+2m-\sqrt{1+4m}}}-\frac{1}{\sqrt{1+2m+\sqrt{1+4m}}},
\Big)=0
\nn
\frac{d^2\lambda_1}{dk_y^2}\Big\vert_{k_y=0}&=&\frac{2+\frac{2}{\sqrt{1+4m}}}{\sqrt{2+4m+2\sqrt{1+4m}}},
\quad
\frac{d^2\lambda_2}{dk_y^2}\Big\vert_{k_y=0}=\frac{2-\frac{2}{\sqrt{1+4m}}}{\sqrt{2+4m-2\sqrt{1+4m}}},
\ee
and
\be
\frac{d^3E}{dk_y^3}\Big\vert_{k_y=0}&=&-(\frac{d^2\lambda_1}{dk_y^2}+\frac{d^2\lambda_2}{dk_y^2})
-3\frac{d\lambda_1}{dk_y}\frac{d^2\lambda_2}{dk_y^2}-3\frac{d^2\lambda_1}{dk_y^2}\frac{d\lambda_2}{dk_y}-\lambda_1\frac{d^3\lambda_2}{dk_y^3}-\lambda_2\frac{d^3\lambda_1}{dk_y^3}
\nn
&=&-\frac{2+\frac{2}{\sqrt{1+4m}}}{\sqrt{2+4m+2\sqrt{1+4m}}}-\frac{2-\frac{2}{\sqrt{1+4m}}}{\sqrt{2+4m-2\sqrt{1+4m}}}=0,
\nn
\frac{d^3\lambda_1}{dk_y^3}\vert_{k_y=0}&=&\frac{3\sqrt{2}}{\sqrt{1+4m}\sqrt{1+2m+\sqrt{1+4m}}}\frac{d\lambda_1}{dk_y}\frac{d^2\lambda_1}{dk_y^2}=0,
\nn
\frac{d^3\lambda_2}{dk_y^3}\vert_{k_y=0}&=&-\frac{3\sqrt{2}}{\sqrt{1+4m}\sqrt{1+2m-\sqrt{1+4m}}}\frac{d\lambda_2}{dk_y}\frac{d^2\lambda_2}{dk_y^2}=0.
\ee
Therefore, near $k_y=0$, we obtain a linear dispersion, $E(k_y)=-k_y+O(k_y^4)$. Based on this information, we can rewrite the corresponding energy eigenstates in this case in the following form:
\be
\psi_{d}/\psi_u&=&i\frac{\lambda_1+k_y}{E+(m-\lambda_1^2+k_y^2)}=i\frac{\sqrt{k_y^2+m+\frac{1}{2}+\frac{1}{2}\sqrt{1+4k_y^2+4m}}+k_y}{-k_y+(m-(k_y^2+m+\frac{1}{2}+\frac{1}{2}\sqrt{1+4k_y^2+4m})+k_y^2)}
\nn
&=&-i+0*\frac{k_y}{m}+0*\frac{k_y^2}{m^2}+0*\frac{k_y^3}{m^3}+O(k_y^4)
\ee
From this expression, it can be seen that the corrections introduced by finite $k_y$ are strongly suppressed by $m$ in the topological insulator. As a result, for an ideal insulator, the $k=0$ approximation for the edge state is sufficiently accurate, and the Newton mass term in the Hamiltonian does not lead to the emergence of quadratic terms that deviate from the linear dispersion. 
Thus, we can use $\psi_0(x)$ as a substitute for $\psi_{k_y}(x)$, even though $e^{-ik_yy}\psi_0(x)$ is not an eigenstate of the Hamiltonian for finite $k_y$.

\end{document}